\documentclass[12pt]{article}

\usepackage{epsfig}
\usepackage{amsmath}
\usepackage{amsthm}
\usepackage{amssymb}
\usepackage{algorithm}
\usepackage{algorithmic}
\usepackage{framed}

\usepackage{url}

\usepackage{fullpage}
\usepackage[utf8x]{inputenc}
\usepackage{amsfonts}
\usepackage{amsmath}
\usepackage{MnSymbol}

\usepackage{graphicx}

\usepackage{natbib}

\usepackage[small,bf]{caption}
\numberwithin{equation}{section}

\newcommand{\bY}{\textbf{Y}}

\newcommand{\bz}{\textbf{z}}
\newcommand{\bZ}{\textbf{Z}}
\newcommand{\bb}{\boldsymbol{\beta}}
\newcommand{\bphi}{\boldsymbol{\phi}}

\newcommand{\bX}{\textbf{X}}
\newcommand{\bF}{\textbf{F}}

\newcommand{\bP}{\textbf{S}}

\newcommand{\bA}{\textbf{A}}
\newcommand{\bB}{\textbf{B}}
\newcommand{\bM}{\textbf{M}}

\newcommand{\bs}{\textbf{s}}
\newcommand{\ba}{\textbf{a}}
\newcommand{\bK}{\textbf{K}}
\newcommand{\br}{\textbf{r}}

\newcommand{\eq}{\begin{equation*}}
\newcommand{\en}{\end{equation*}}
\newcommand{\eqa}{\begin{eqnarray*}}
\newcommand{\ena}{\end{eqnarray*}}
\newcommand{\eqn}{\begin{equation}}
\newcommand{\enn}{\end{equation}}

\newcommand{\be}{\begin{equation}}
\newcommand{\ee}{\end{equation}}

\newcommand{\eqan}{\begin{eqnarray}}
\newcommand{\enan}{\end{eqnarray}}

\newcommand{\vp}{\varpropto}

\newcommand{\pmat}{\begin{pmatrix}}
\newcommand{\pman}{\end{pmatrix}}

\long\def\symbolfootnote[#1]#2
   {\begingroup\def\thefootnote{\fnsymbol{footnote}}\footnote[#1]{#2}\endgroup}

\graphicspath{{../figures/}}

\begin{document}

\def\spacingset#1{\renewcommand{\baselinestretch}%
{#1}\small\normalsize} %\spacingset{1.5}

\title{
Exact Hamiltonian Monte Carlo  for
\\ Truncated  Multivariate Gaussians 
}

\author{Ari Pakman\thanks{ari@stat.columbia.edu} \, and Liam Paninski\thanks{liam@stat.columbia.edu} 
\\
\\ 
Department of Statistics,  
\\Center for Theoretical Neuroscience 
\\
and Grossman Center for the Statistics of Mind
\\
\\
Columbia University
}
\date{}
\maketitle

\abstract{We present a Hamiltonian Monte Carlo algorithm to sample
from multivariate Gaussian distributions in which the target space is
constrained by linear and quadratic inequalities or products thereof.
The Hamiltonian equations of motion can be integrated exactly and
there are no parameters to tune. The algorithm mixes faster and is more efficient 
than Gibbs sampling.   The runtime depends on the number and shape of the constraints 
 but the algorithm is highly
parallelizable.  
In many cases, we can exploit special structure in the covariance matrices of the untruncated Gaussian to further speed 
up the runtime. A simple extension of the algorithm permits sampling
from distributions whose log-density is piecewise quadratic, as in the
``Bayesian Lasso'' model.  }

\thispagestyle{empty}

\newpage
\pagestyle{plain}
\setcounter{page}{1}

\section{Introduction}

The advent of Markov Chain Monte Carlo methods  has made it possible to sample from complex multivariate probability distributions \citep{robert2004monte},
 leading to a remarkable progress in Bayesian modeling, with applications to many areas of applied statistics and machine 
learning \citep{gelman2004bayesian}. 

In many cases, the data or the parameter space are constrained~\citep{gelfand1992bayesian} and 
the need arises for efficient sampling techniques for truncated distributions. 
In this paper we will focus on the Truncated Multivariate Gaussian (TMG),  a  
$d$-dimensional  multivariate Gaussian distribution of the form  
\eqan
\log p(\bX) = -\frac12 \bX^T \bM  \bX + \br^T  \bX + const.
\label{gaussiani}
\enan
with $\bX, \, \br \in \mathbb{R}^d$ and $\bM$ positive definite, subject to $m$ inequalities
\eqan
Q_j(\bX)  \geq 0 \qquad j=1, \ldots,m\,,
%\bF_j \cdot \bX + g_j \geq 0 \qquad j=1, \ldots,m\,.
\label{ineqi}
\enan where $Q_j(\bX)$ is a product of linear and quadratic
polynomials.  These distributions play a central role in models as
diverse as the Probit and Tobit
models~\citep{albert1993bayesian,tobin1958estimation}, the
dichotomized Gaussian model~\citep{emrich1991method, cox2002some},
stochastic integrate-and-fire neural
models~\citep{paninski2004maximum}, Bayesian isotonic
regression~\citep{neelon2004bayesian}, the Bayesian bridge model
expressed as a mixture of Bartlett-Fejer
kernels~\citep{polson2011bayesian}, and many others.

\begin{figure}[t]
\begin{center}
\includegraphics[scale=.87]{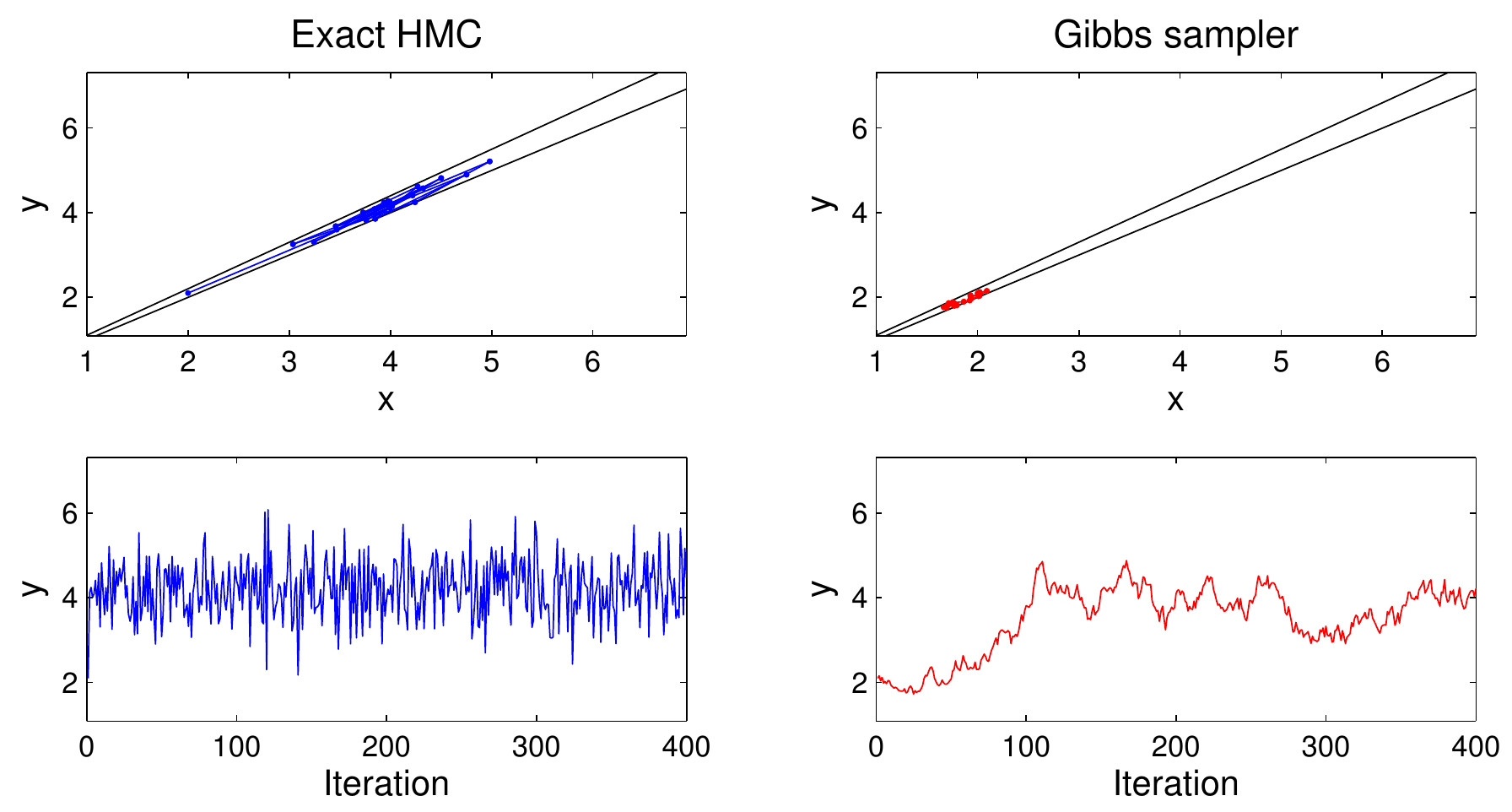}
\end{center}
\caption{{\bf HMC vs Gibbs sampler.} 
\emph{Comparison for a two-dimensional distribution with $\log p(x,y) \vp -\frac12 (x-4)^2  -\frac12 (y-4)^2$, constrained to the wedge $x \leq y \leq 1.1x$ and $x, y \geq 0$.
The initial point is $(x,y)=(2,2.1)$. 
Upper panels: first 20 iterations. Lower panels: Second coordinate of the first 400 iterations.
In the HMC case we used $T=\pi/2$. The HMC sampler moves rapidly to oscillate around~$y=4$, as desired, while the Gibbs sampler mixes relatively slowly. }}
\label{hmc_vs_gibbs}
\end{figure}

% which is  of interest both by itself 
% and  as a building block in more complex models that can be expressed as mixtures of Gaussians (e.g. the Bayessian Lasso, t-distributions).
The standard approach to sample from TMGs  is the Gibbs sampler~\citep{geweke1991efficient, kotecha1999gibbs}.
The latter reduces the problem to one-dimensional truncated Gaussians, for which simple and  efficient sampling methods exist~\citep{robert1995simulation, damien2001sampling}.
While it enjoys the benefit of having no parameters to tune, the Gibbs sampler can suffer from two problems, which make it inefficient in some  situations. 
Firstly, its runtime scales linearly with the number of dimensions. 
Secondly, even though a change of variables that maps $\bM$ in (\ref{gaussiani}) to the identity often  improves the mixing speed~\citep{rodriguez2004efficient},
the exploration of the target space can still be very slow when the constraints  (\ref{ineqi}) impose high correlations among the coordinates. 
Figure~\ref{hmc_vs_gibbs} illustrates this effect in a simple example.
Improvement over the Gibbs runtime can be obtained with a hit-and-run algorithm \citep{chen1992application}, but the latter suffers from the same slow convergence problem
when the constraints impose strong correlations.

In this paper we present an alternative algorithm to sample from TMG distributions for constraints~$Q_j(\bX)$ in~(\ref{ineqi}) given by linear or quadratic 
functions or products thereof, based on the
Hamiltonian Monte Carlo (HMC) approach. The HMC method, introduced in \cite{duane1987hybrid}, considers
the log of the probability distribution as minus the potential energy of a particle, and introduces a Gaussian distribution for momentum variables in order to define a 
Hamiltonian function. The method generally avoids random walks and mixes faster than  Gibbs or Metropolis-Hastings techniques.
The HMC sampling procedure alternates between sampling the Gaussian momenta and letting 
the position of the particle evolve by integrating its Hamiltonian equations of motion. 
In most models, the latter cannot be integrated exactly, so the 
resulting position is used as a Metropolis proposal, with an acceptance probability that depends exponentially
on the energy gained due to the numerical error. The downside is that two parameters 
must be fine-tuned for the algorithm to work properly: the integration time-step size and the number of time-steps.
In general the values selected correspond to a compromise between a high acceptance rate and a good rate of exploration of the space~\citep{hoffman2011no}.
More details of HMC  can be found in the reviews by~\cite{kennedy1990theory} and~\cite{neal2010mcmc}. 

The case  we consider in this work is special because the Hamiltonian equations of motion can be integrated exactly, 
thus leading to the best of both worlds:  HMC mixes fast and, as in Gibbs, there are no parameters to tune
and the Metropolis step always accepts (because the energy is conserved exactly).
The truncations~(\ref{ineqi})  are incorporated via hard walls, against which the particle bounces off elastically.
The runtime depends highly on the shape and location of the truncation, as most of the computing time goes into finding the time of the next wall bounce
and the direction of the reflected particle. But unlike the Gibbs sampler, these computations are parallelizable,  potentially allowing 
fast implementations.

The idea behind the HMC sampler for a Gaussian in a truncated space turns out to be 
applicable also when the log-density is piecewise quadratic.
We show that a simple extension of the algorithm allows us to sample
from such distributions, focusing on the example of the ``Bayesian
Lasso'' model~\citep{park2008bayesian}.

Previous HMC applications that  made use of  exactly solvable Hamiltonian equations include 
sampling from non-trivial 
integrable Hamiltonians~\citep{kennedy1994exact},  and  
importance sampling, with
the target distribution approximated by a distribution with  an integrable Hamiltonian~\citep{rasmussen2003gaussian, izaguirre2004shadow}.

In the next Section we present the new sampling algorithm for linear
and quadratic constraints. In Section~\ref{examples} we present four  example applications.
In all our examples, the matrix $\bM$ or its inverse have a special structure that allows us to accelerate the runtime of the sampler.
In Section~\ref{baylasso} we discuss the extension to the Bayesian Lasso
model. We have implemented the sampling algorithm in the R package ``tmg,'' available in the CRAN repository.

\section{The Sampling Algorithm}
\subsection{Linear Inequalities}
\label{Linear}
Consider first sampling from
\eqan
\log p(\bX) = -\frac12 \bX \cdot \bX + const.
\label{gaussian}
\enan
subject to
\eqan
\bF_j \cdot \bX + g_j \geq 0 \qquad j=1, \ldots,m\,.
\label{ineq}
\enan
Any quadratic form for $\log p(\bX)$, as in (\ref{gaussiani}), can be brought to the above canonical form by a linear change of variables.
Let us denote the components of $\bX$ and $\bF_j$  as
\eqan
\bX &=& (x_1, \ldots, x_d) \,,
\\
\bF_j &=& (f_j^1, \ldots, f_j^d) \,.
\enan 
In order to apply the HMC method, we introduce momentum variables $\bP$,
\eqan
\bP = (s^1, \ldots, s^d) \,,
\enan
and consider the Hamiltonian
\eqan
H = \frac12 \bX \cdot \bX + \frac12 \bP \cdot \bP \,,
\label{H}
\enan
such that the joint distribution is $ p(\bX, \bP) = \exp(-H)$.
The equations of motion following from~(\ref{H}) are
\eqan
\dot{x}_i &=& \frac{\partial H}{\partial s^i} = s^{i}
\\
\dot{s}^i &=& -\frac{\partial H}{\partial x_i} = - x_{i}  \qquad \qquad i = 1 , \ldots, d
\enan
which can be combined to 
\eqan
\ddot{x}_i = - x_i  \,,
\label{eom}
\enan
and have a solution
\eqan
x_i(t) &=& a_i \sin(t) + b_i \cos(t) \,.
\label{sol1}
\enan
The constants  $a_i, b_i$ can be expressed in terms of the initial conditions as
\eqan
a_i &=& \dot{x}_i(0) = s^i(0)
\label{ai}
\\
b_i &=& x_i(0)
\enan The HMC algorithm proceeds by alternating between two steps. In
the first step we sample $\bP$ from $p(\bP) = {\cal N}(0,\mathbb{I}_d)$. 
In the second step we use this $\bP$ and the last
value of $\bX$ as initial conditions, and let the particle move during
a time $T$, after which the position and momentum have values~$\bX^*$ and~$\bP^*$.
The value $\bX^*$ belongs
to a Markov chain with equilibrium distribution $p(\bX)$.  
To see this, note that for a given $T$, the particle trajectory is deterministic once the momentum $\bP$ is sampled,
so the transition probability is 
\eqan 
p(\bX^*|\bX,T) = p(\bP)   \left|  \partial \bP/ \partial  \bX^*   \right|   \,.
\label{trans}
\enan 
Two important properties of Hamiltonian dynamics are the conservation of energy
\eqan 
H(\bX, \bP) = H(\bX^*, \bP^*)
\enan 
and the conservation of volume in phase space
\eqan 
d \bP^*  d \bX^* = d \bP d \bX
\enan 
which implies the equation
\eqan 
\left|  \partial \bP/ \partial  \bX^*   \right| = \left|  \partial \bP^*/ \partial  \bX   \right| \,.
\enan 
From the above results the detailed balanced condition follows as
\eqan
p(\bX)p(\bX^*|\bX,T)  &=& p(\bX)p(\bP)  \left|  \partial \bP/ \partial  \bX^*   \right| 
\\
&=& e^{-H(\bX, \bP  )}   \left|  \partial \bP/ \partial  \bX^*   \right| 
\\
&=&  e^{-H(\bX^*, -\bP^*  )}    \left|  \partial \bP^*/ \partial  \bX   \right| 
\\
&=& p(\bX^* )p(\bX|\bX^*, T)
\enan 
where  we used the invariance of $H$ under  $\bP \rightarrow -\bP$. We will discuss the appropriate choice for~$T$ in Section~\ref{efficiency}.

The trajectory of the particle is given by (\ref{sol1}) until it hits a wall, and this occurs 
whenever any of the inequalities~(\ref{ineq}) is saturated. To find the time at which this occurs, it is convenient to define 
\eqan
K_j(t) &=& \sum_{i=1}^{d}f_j^ix_i(t) + g_j  \qquad j=1, \ldots,m\,.
\\
&=&  \sum_{i=1}^{d}f_j^ia_i \sin(t) + \sum_{i=1}^{d}f_j^ib_i \cos(t) + g_j
\\
&=& u_j \cos(t + \varphi_j) + g_j
\label{ug}
\enan
where
\eqan
u_j &=& \sqrt{ (\sum_{i=1}^d f_j^i a_i)^2 +    (\sum_{i=1}^d f_j^i b_i)^2 } \,,
\label{udef}
\\
\tan \varphi_j &=& -\frac{\sum_{i=1}^d f_j^i a_i}{\sum_{i=1}^d f_j^i b_i} \,.
\label{phidef}
\enan
Along the trajectory we have $K_j(t) \geq 0$ for all $j$ and  
a wall hit corresponds to $K_j(t)=0$, so from (\ref{ug}) it follows that the particle can only reach those walls 
satisfying $u_j > |g_j|$. Each one of those reachable walls has associated two times $t_j>0$ such that
\eqan
K_j(t_j) =0 \,,
\label{tdef}
\enan
and the actual wall hit corresponds to the smallest of all these times.
%Some computational time can be saved by considering first those constraints $j$ for $K_j(0)>0$ and $\dot{K}_j(0)<0$,
%or $K_j(0)<0$ and $\dot{K}_j(0)>0$, since they will reach $K_j(t)=0$ first.
Suppose that the latter occurs for $j=h$. 
At the hitting point, the  particle bounces off the wall and the trajectory continues with a reflected velocity. The latter can be obtained by noting that the 
vector $\bF_h$ is perpendicular to the reflecting plane. Let us decompose the  velocity  as
\eqan
\dot{\bX}(t_h) = \dot{\bX}_{\perp}(t_h) +  \alpha_h  \bF_h \,,
\label{dec_vel}
\enan
where  %$\langle  \bF_j, \dot{\bX}_{\perp}(t_j) \rangle=0 $ 
$\bF_h \cdot \dot{\bX}_{\perp}(t_h) =0 $ 
and  
\eqan
%\alpha_j &=& \frac{\langle \dot{\bX}(t_j) ,\bF_j \rangle}{||\bF_j||^2} \,.
\alpha_h &=& \frac{  \bF_h \cdot \dot{\bX}(t_h) }{||\bF_h||^2} \,.
\enan
The reflected velocity, $\dot{\bX}_{R}(t_h) $,  is obtained  by inverting  the component perpendicular to the reflecting plane
\eqan
\dot{\bX}_{R}(t_h) 
&=&  \dot{\bX}_{\perp}(t_h)  -\alpha_h  \bF_h  \,,
\\
&=&  \dot{\bX}(t_h) - 2 \alpha_h \bF_h \,.
\label{refl_vel}
\enan
It is easy to verify that this transformation leaves the Hamiltonian (\ref{H}) invariant.
Once the reflected velocity is computed, we use it as an initial condition in (\ref{ai}) to continue the particle trajectory.

If we prefer to keep the original distribution in
the form (\ref{gaussiani}), we should consider the Hamiltonian 
\eqan H =
\frac12 \bX^T \bM \bX - \br^T \bX + \frac12\bP^T \bM^{-1} \bP\,.
\label{hamnw}
\enan 
This election for the mass matrix leads to the simple solution
\eqan
x_i(t) &=& \mu_i + a_i \sin(t) + b_i \cos(t) \,,
\label{solm}
\enan
where 
\eqan 
\mu_i &=& \sum_{j=1}^{d} M_{i j}^{-1}r_j \,,
\\
a_i &=&  \dot{x}_i(0) = \sum_{j=1}^{d} M_{i j}^{-1}s_j(0) \,,
\\
b_i &=& x_i(0) -\mu_i \,.
\enan
Since the particle trajectory depends on $\dot{x}_i(0)$, we can start each iteration by sampling 
$\dot{x}_i(0) \sim {\cal N}(0,\bM^{-1})$, instead of sampling $s_i(0) \sim {\cal N}(0,\bM)$ itself.

\subsection{The runtime of the sampler}
\label{runtime}
The runtime of each iteration has a contribution that scales linearly with $m$, the number constraints, since we have
to compute the $m$ values~$u_j$ at (\ref{udef}).
But the dominant computational time goes to 
compute $\varphi_j$ and $t_j$, defined in (\ref{phidef}) and
(\ref{tdef}), which only needs to be done when 
$u_j > |g_j|$. The average number of coordinates for which this condition occurs, as well 
as the number of times the particle hits the walls per iteration, 
varies according to the value of $T$ and the shape and location of the walls.

The  sums in expressions (\ref{udef}) and~(\ref{phidef}) can be interpreted as
matrix-vector multiplications, with cost $O(md)$ for general
constraint matrices $\bF = (\bF_1^T \bF_2^T \ldots \bF_m^T)^T$.  Note
that these matrix-vector multiplications are highly parallelizable.
In addition, in many cases there may be some special structure that
can be exploited to speed computation further; for example, if $\bF$
can be expressed as a sparse matrix in a convenient basis, this cost
can be reduced to $O(d)$.  

Also, in both frames (\ref{H}) and (\ref{hamnw})  we must act, 
for each sample, with a matrix $R^{-1}$, where $\bM=R^TR$. 
Equivalently, we must multiply by $Z^T$, where $\bM^{-1}=\Sigma=Z^TZ$.
In the frame~(\ref{H}) this is needed because the  samples must be mapped back to the original frame~(\ref{gaussiani}) as
\eqan
\bX \rightarrow R^{-1}\bX + \bM^{-1}\br\,,
\enan 
and in the frame~(\ref{hamnw}) the action of $R^{-1}$ is needed in order to sample the initial velocity from ${\cal N}(0,\bM^{-1})$ at each iteration.
The action of $R^{-1}$ takes $O(d^2)$ generally, but in some cases the matrix~$\bM$ or its inverse have a special structure that allows  us to accelerate this operation.
This is the case in the four example applications  we present in Section~\ref{examples}.

Regarding which of the two frames (\ref{H}) and (\ref{hamnw}) is preferred, this depends on the nature of the constraints,
since the latter change when we transform the quadratic form for $\log p(\bX)$ in~(\ref{gaussiani})  to the
form (\ref{gaussian}). For example, a sparse set of constraints $\bF$ in the original frame leads to a fast evaluation of 
(\ref{udef}) and~(\ref{phidef}), but the  wall geometry in the transformed frame 
may lead to a smaller number of wall hits and therefore shorter runtime.

\subsection{Quadratic and Higher Order Inequalities}
\label{quadratic}

The sampling algorithm can be extended in principle  to polynomial constraints of the form
\eqan
Q_j(\bX)  \geq 0 \qquad j=1,\ldots, m\,.
\label{pol}
\enan
Evaluating $Q_j(\bX)$ at the solution (\ref{sol1}) leads to a polynomial in $\sin(t)$ and $\cos(t)$, whose zeros must be found in order to 
find the hitting times.  When a wall is hit, we reflect the velocity by inverting the sign of the component perpendicular to the wall, given by
the gradient ${\bf \nabla} Q_j(\bX)$. This vector plays a role similar to~$\bF_j$ in (\ref{dec_vel})-(\ref{refl_vel}).
Of course, for general polynomials $Q_j(\bX)$ computing the hitting times might be numerically challenging.

\begin{figure}[t]
\begin{center}
\includegraphics[scale=.64]{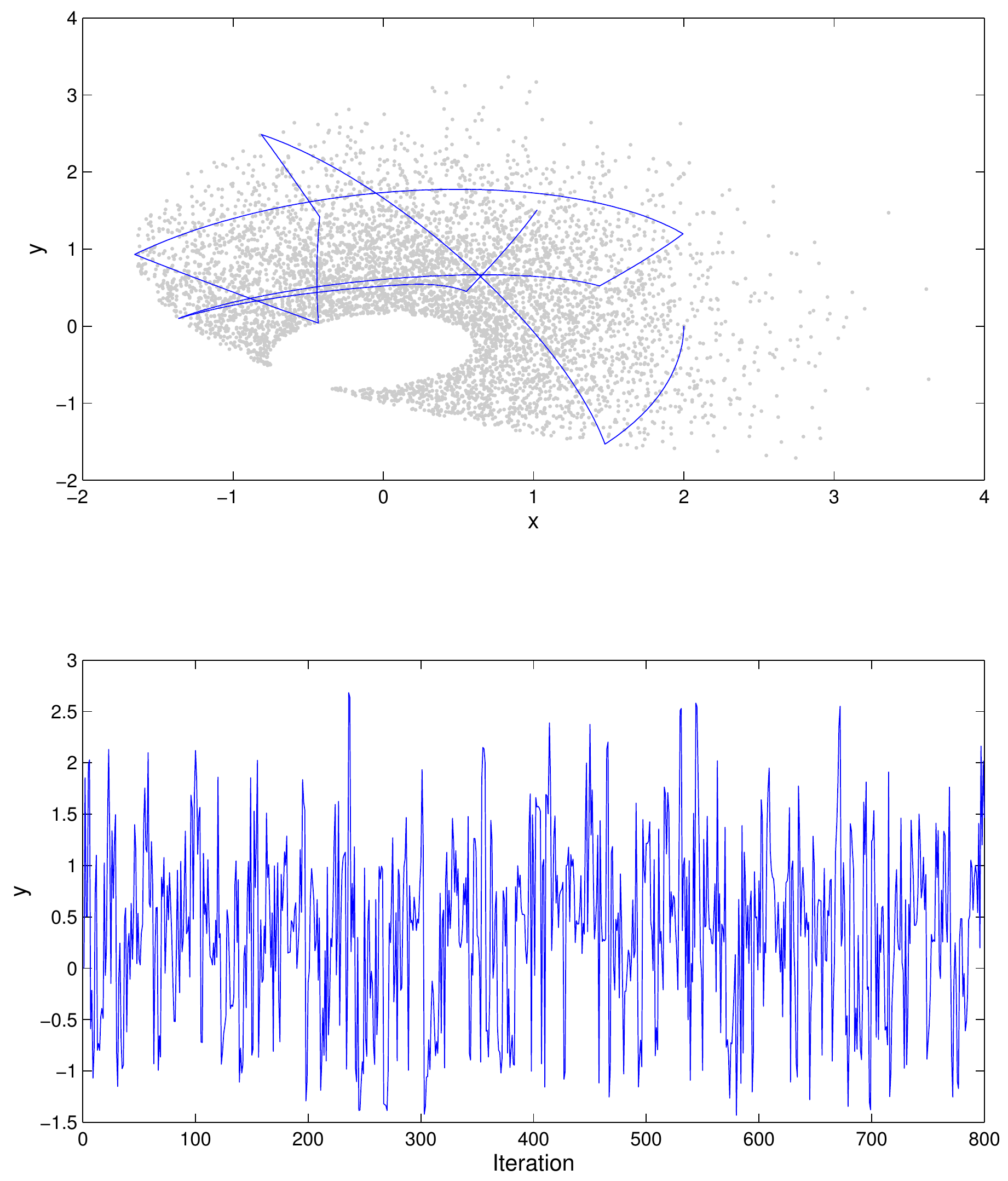}
\end{center}
\caption{{\bf Truncation by quadratic inequalities.} 
\emph{Above: 6000 samples of a two-dimensional canonical normal distribution, constrained by the quadratic inequalities (\ref{qq1}) -(\ref{qq2}). The piecewise elliptic curve
shows the trajectory of the particle in the first iterations, with starting point $(x,y)= (2,0)$. Below: first $800$ iterations of the vertical coordinate.
For the algebraic solution of (\ref{quart}), we used the C++ code from the DynamO package~\citep{JCC:JCC21915}.
}}
\label{ellipses_fig}
\end{figure}

One family of solvable constraints involves  quadratic inequalities of the form
\eqan
Q_j(\bX)=\bX^T \bA_j \bX + \bX \cdot \bB_j + C_j \geq 0 \qquad j=1,\ldots, m\,,
\label{ellipse}
\enan
where $\bA_j \in \mathbb{R}^{d,d}, \, \bB_j \in \mathbb{R}^d,\,  C_j \in \mathbb{R}$.
For statistics applications where these constraints are important, see e.g.~\cite{ellis2007multivariate}.
Inserting (\ref{sol1}) in the equality for (\ref{ellipse}) leads, for each $j$,  to the following equation for the hitting time:
\eqan
q_1 \cos^2 (t) + q_2 \cos (t) + q_3 = - \sin (t) (q_4 \cos (t) + q_5) \,,
\label{quadcs}
\enan
with
\eqan
q_1 &= & \sum_{i,k} A^{ik}b_i b_k -\sum_{ik} A^{ik}a_ia_k \,,
\\
q_2 &= &  \sum_{i} B^i b_i \,,
\\
q_3 &= & C + \sum_{ik} A^{ik}a_ia_k \,,
\\
q_4 &= & 2 \sum_{i,k} A^{ik}a_ib_k \,,
\\
q_5 &= & \sum_{i} B^i a_i \,,
\enan
and we omitted the $j$ dependence to simplify the notation.
If the ellipse in (\ref{ellipse}) is centered at the origin, we have $\bB_j=q_2=q_5=0$, and equation (\ref{quadcs})
simplifies to
\eqan
q_1 + 2q_3 + u \sin(2t+ \varphi) = 0
\label{wb}
\enan
where
\eqan 
u^2 &=& q_1^2 + q_4^2 \,,
\\
\tan \varphi &=& \frac{q_1}{q_4} \,,
\enan 
and the hit time can be found from (\ref{wb}) as in the linear case. 
In the general  $\bB_j \neq 0$ case,  the square of (\ref{quadcs}) gives the quartic equation
\eqan
r_4 \cos^4 (t) + r_3 \cos^3 (t)  + r_2 \cos^2 (t) + r_1 \cos (t) + r_0=0 \,,
\label{quart}
\enan
where
\eqan 
r_4 &=& q_1^2 + q_4^2 \,,
\\
r_3 &=& 2q_1 q_2 + 2 q_4 q_5 \,,
\\
r_2 &=& q_2^2 + 2q_1q_3 + q_5^2 - q_4^2 \,,
\\
r_1 &=& 2q_2 q_3 - 2q_4q_5 \,,
\\
r_0 &=& q_3^2 - q_5^2 \,.
\enan 
Equation (\ref{quart}) can be solved exactly for $cos(t)$ using standard algebraic methods~\citep{herbison1994solving}.
A wall hit corresponds, among all the constraints $j$, to the solution for $\cos(t)$ with smallest  $t>0$  and $|\cos(t)| \leq 1$,  which also solves~(\ref{quadcs}). 
As an example, Figure~\ref{ellipses_fig} shows samples from a two-dimensional canonical normal distribution, constrained by 
\eqan
\frac{(x-4)^2}{32} + \frac{(y-1)^2}{8} &\leq& 1 \,,
\label{qq1}
\\
4x^2 + 8y^2 -2xy + 5y &\geq& 1 \,.
\label{qq2}
\enan
Equipped with the  results for linear and quadratic constraints, we can also find the  hitting times for constraints of the form
\eqan
Q(\bX) = \prod_{j}Q_j(\bX) \geq 0
\label{decomp}
\enan
where each $Q_j(\bX)$ is a linear or a quadratic function.
Each factor defines an equation  as (\ref{tdef}) or (\ref{quart}), and the hitting time is the smallest at which any factor becomes zero.
For other polynomials, one has to resort to numerical methods to find the hitting times.

\subsection{Travel time and efficiency}
\label{efficiency}
The value of $T$ should be chosen to make the sampling as efficient as possible. This is not easy 
because the path of the particle will be determined by the wall bounces, so it is  difficult to compute in advance.
A safe strategy is to choose a value of $T$ that makes the sampling efficient at least for those trajectories with no wall hits.
The efficiency can be quantified via the Effective Sample Factor (ESF) and Effective Sample Size (ESS)~\citep{liumonte}.
Let us call  the samples  $\bX^{(p)}$.  The variance in the estimation of the expected value of a function $h(\bX)$ using $m$ samples is
\eqan 
var\left( \frac{h(\bX^{(1)}) + \cdots + h(\bX^{(m)})   }{m} \right) = \frac{var(h(\bX))}{m}  \left[ 1 + \sum_{j=1}^{m-1} 2\left(1-\frac{j}{m}  \right) \rho_j \right]
\label{hvar}
\enan 
with  $\rho_j$ the autocorrelation function (ACF),
\eqan 
\rho_j &=& corr[ h(\bX^{(1)}), h(\bX^{(1+j)})     ]  \,.
\enan 
The ESF and ESS are defined as 
\eqan 
\textrm{ESF} &=&  \left[ 1 + \sum_{j=1}^{m-1} 2 \left(1-\frac{j}{m}  \right) \rho_j \right]^{-1} \,,
\label{ESF}
\\
\textrm{ESS} &=& m \times \textrm{ESF } \,,
\enan 
and a sampling scheme is more efficient when its ESF is higher, because that leads  to a lower variance in (\ref{hvar}).

For concreteness, let us consider the estimation of the mean of a coordinate $x_i$ in the frame in which the Hamiltonian is given by~(\ref{H}).
In the absence of wall hits, from (\ref{sol1})-(\ref{ai}) it follows that  successive samples are given by
\eqan 
x^{(p+1)}_i = x_i^{(p)} \cos(T) + s_i^{(p)} \sin(T) \,.
\enan 
Assuming that the momenta samples are i.i.d., i.e.,
\eqan 
\langle s_i^{(p)} s_i^{(q)}  \rangle = \delta_{pq} \,,
\label{iid}
\enan 
it is easy to see that the ACF is given by
\eqan 
\rho_j = \langle x_i^{(1)} x_i^{(j+1)} \rangle =  \cos^j(T)\,.
\enan 
In principle we could plug this expression into (\ref{ESF}) and find the value of $T$ that maximizes the ESF. In particular, 
when $\cos(T)<0$, there are values of $T$ which lead to a super-efficient sampler with ESF $>1$, so that the variance~(\ref{hvar})  
is  {\emph{smaller}} than using  i.i.d.~samples (i.e., with  $\rho_j=0$).
This is similar to the antithetic variates method to reduce the variance of Monte Carlo estimates~\citep{hammersley1956new}.

\begin{table}[t]
\begin{center}
\begin{tabular}{|l|c|l|c|}
  \hline
   & Avg. CPU Time &  & $y$
  \\
  \hline \hline 
  $\textrm{HMC}_{T=\pi/2}$ & 3.78 secs &  ESF  &  { \bf 2.7 (1.18/5.11)  } \\
  \cline{3-4}
    &      & ESS/CPU   & {\bf 5,802 (2,507/10,096)  } \\
          
  \hline \hline 
  $\textrm{HMC}_{T=\pi/10}$ & 1.38 secs &  ESF    &   0.052 (0.033/0.094)    \\
  \cline{3-4}
     &     & ESS/CPU   & 301 (191/547) \\        
          
  \hline \hline 
  Gibbs &  0.57 secs & ESF  & 0.017 (0.011/0.032) \\
    \cline{3-4}
     &  &  ESS/CPU    & 237 (154/456)  \\  
  \hline
\end{tabular}
\end{center}
\caption{{\bf HMC vs Gibbs sampler.}  
\emph{Comparison of efficiency criteria for 30 runs of three  samplers in the example of Figure~\ref{hmc_vs_gibbs}.
For the definition of ESF and ESS, see Section~\ref{efficiency}. ESS/CPU is the Effective Sample Size in units of the CPU runtime.
A sampler is more efficient for higher ESF and ESS/CPU.
In each run,  we used 8,000 samples, after discarding  2,000 samples as burn-in.
The CPU time for each run is random, but its variability across runs is negligible due to the large number of samples.
For ESF and  ESS/CPU, we show median (first/third quartiles).
For the Gibbs sampler,  we used the algorithm of~\citep{damien2001sampling}.
The runtime for HMC is relatively high due to the many  reflections of the particle against the walls. 
Note that the efficiency of the HMC sampler depends strongly on the value of the travel time $T$. }}
\label{table_ess_wedge}
\end{table}

In practice, we have found that the above strategy is not very useful. The assumption~(\ref{iid}) holds only approximately 
for standard generators of random variables, and this leads to very unstable values for the ESF for any value of $T$. 
A safe choice is to use $T=\pi/2$, which leads to 
\eqan 
x^{(p+1)}_i = s_i^{(p)} \,,
\label{xep}
\enan 
so our sampling of  the space $\bX$, for trajectories with no wall hits, will be as efficient as our method to sample the momenta $\bP$.
Since the i.i.d. assumption holds approximately, this leads to values of ESF that bounce around $1$.  The ESF in a general case will also depend on the shape and location
of the constraint walls. But we have found that the  super-efficient case ESF $ > 1$ is not uncommon, although as mentioned, the actual value of the ESF is quite unstable.
Table~\ref{table_ess_wedge} illustrates the big difference in the efficiency of the HMC sampler for $T=\pi/2$ and $T=\pi/10$ in the two-dimensional 
example of Figure~\ref{hmc_vs_gibbs}. A similar reasoning suggests the use of  $T=\pi/2$ also when working in the frame in which the Hamiltonian is given by (\ref{hamnw}).
In the next two Sections we adopted~$T=\pi/2$, and we rotated the coordinates to a canonical frame with Hamiltonian~(\ref{H}).

In Tables~\ref{table_ess_wedge} and~\ref{table_ess}  we compare the efficiency of our HMC method 
to the Gibbs sampler. To implement the latter efficiently  we made two choices.
Firstly, we rotated the coordinates to a canonical frame in which the unconstrained Gaussian has unit covariance. This transformation 
often makes the Gibbs sampler mix faster~\citep{rodriguez2004efficient}. For the Gibbs sampler itself, we used 
the slice sampling version of~\citep{damien2001sampling}. This algorithm augments by one the number of variables, but turns the conditional distributions 
of the coordinates of interest into uniform distributions. The latter are much faster to sample than the truncated one-dimensional Gaussians one gets  otherwise.
We  checked that the efficiency of the algorithm of~\citep{damien2001sampling},   measured in units of ESS/CPU time, 
is much higher  than that of the Gibbs sampler based on direct sampling from
the one-dimensional truncated normal using inverse cdf or rejection
sampling.

\section{Examples}
\label{examples}
In this Section we present four example applications of our algorithm. In the first example, we present a detailed efficiency comparison between the HMC and the Gibbs samplers.
As mentioned in Section~\ref{runtime}, in both frames (\ref{H}) and (\ref{hamnw}), for each sample of the HMC we must act with a matrix $R^{-1}$, where $\bM=R^TR$, or multiply
by $Z^T$, where $\bM^{-1}=\Sigma=Z^TZ$. In all our examples, we show how some special structure of  $\bM$ or $\Sigma$ allows us to accelerate these operations.

\subsection{Probit and Tobit Models}
\label{example1}
The Probit model is a popular discriminative probabilistic model for binary classification with continuous inputs~\citep{albert1993bayesian}.
The conditional probabilities for the binary labels $y = \pm 1$ are given by
\eqan
p(y = -1 |\bz, \bb) &=& \Phi(\bz \cdot \bb) = \frac{1}{\sqrt{2 \pi}}\int_{- \infty}^{\bz \cdot \bb} \!\!\!\!\!\! dw \, e^{- \frac{w^2}{2}}
\\
&=& \frac{1}{\sqrt{2 \pi}}\int_{- \infty}^{0}  \!\!\!\!\!\! dw \, e^{- \frac{(w+ \bz \cdot \bb)^2}{2}}
\\
p(y = +1 |\bz, \bb) &=& 1-\Phi(\bz \cdot \bb)
\\
&=& \frac{1}{\sqrt{2 \pi}}\int^{+ \infty}_{0} \!\!\!\!\!\! dw \, e^{- \frac{(w+ \bz \cdot \bb)^2}{2}}
\enan
where  $\bz \in \mathbb{R}^{p}$ is a vector of regressors and $\bb \in \mathbb{R}^{p}$ are the parameters of the model.
Given $N$ pairs of labels and regressors
\eqan
\textbf{Y} = (y_1, \ldots, y_N) \,,
\\
\textbf{Z} = (\bz_1, \ldots, \bz_N) \,,
\enan
the posterior distribution of the parameters $\bb$  is 
\eqan
p(\bb| \textbf{Y}, \textbf{Z}) &\vp& p(\bb) \prod_{i=1}^{N} p(y_i|\bz_i, \bb)
\\
&\vp&  p(\bb)  \int_{y_i w_i \geq 0} \!\!\!\!\!\!\!\!\!\!\!\! dw_1 \ldots dw_N \,\, e^{-\frac12 \sum_{i=1}^N (w_i+ \bz_i \cdot \bb)^2}  \qquad i=1,\ldots, N,
\label{post}
\enan
where $p(\bb)$ is the prior distribution. 
\begin{figure}[t]
\begin{center}
\includegraphics[scale=.90]{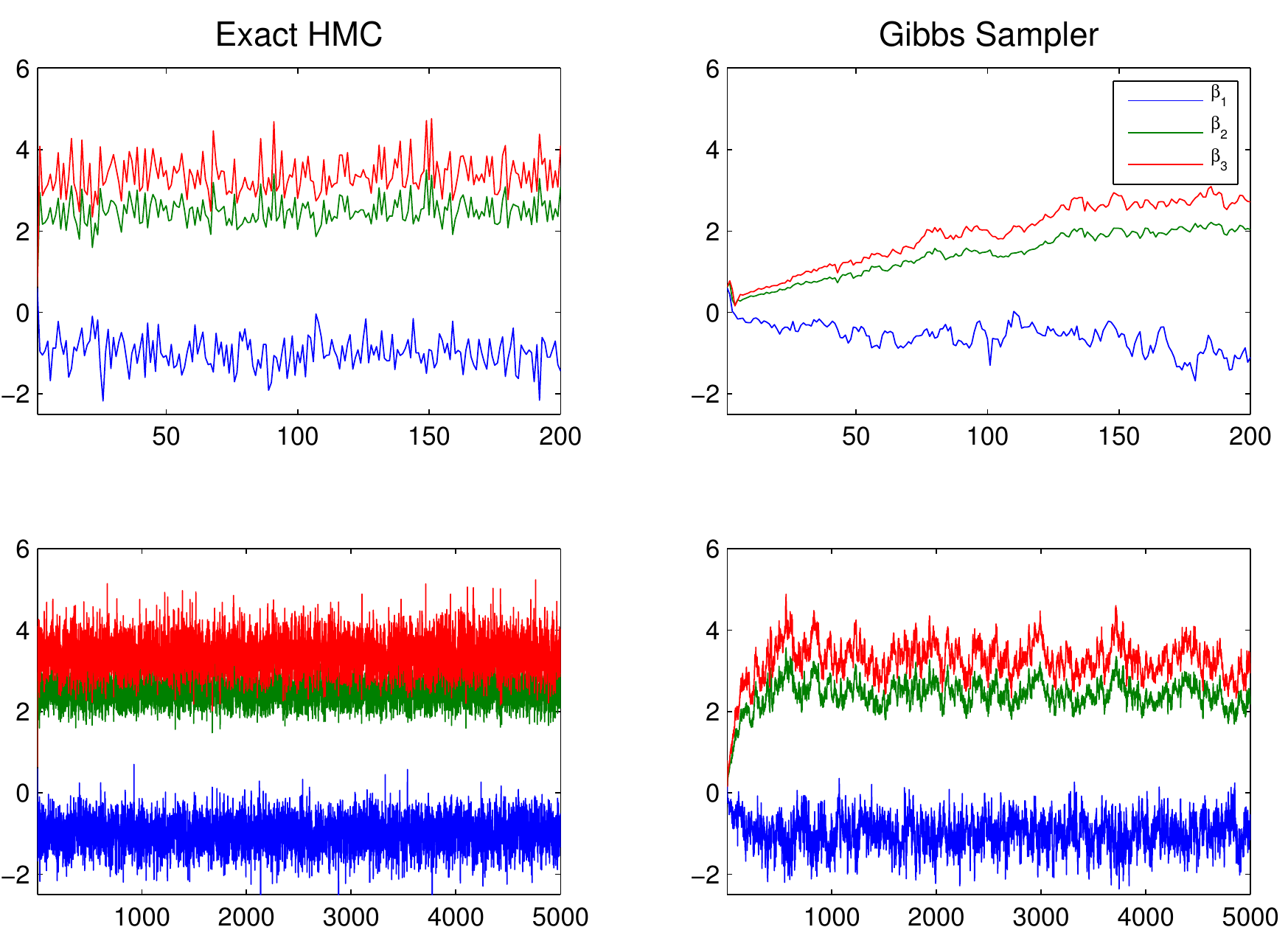}
\end{center}
\caption{{\bf Bayesian Probit model.} 
\emph{First 200 and 5000 samples from the posterior (\ref{post}) of a model with $p=3$.  $N=800$ pairs $(y_i,z_i)$ 
were generated with $\beta_1=-9, \beta_2=20, \beta_3=27$ and we assumed a Gaussian prior with zero mean and $\sigma^2=1$.  
Note that the means of the sampled values are different from the values used to generate the data, due to the zero-mean prior.  Left: Exact HMC sampler. Right: Gibbs sampler, 
with whitened covariance to improve mixing~\citep{rodriguez2004efficient}.}}
\label{probit_fig}
\end{figure}
The likelihood $p(y_i|\bz_i, \bb)$ corresponds to a  model
\eqan
y_i &=& sign(w_i)
\\
w_i &=& -\bz_i \cdot \bb + \varepsilon_i
\\
\varepsilon_i &\sim& {\cal N}(0,1)
\enan
in which only the sign of $w_i$ is observed, but not its value.
Assuming a Gaussian prior with zero mean and covariance $\sigma^2 \mathbb{I}_p$, 
expression (\ref{post}) is the marginal 
distribution of a multivariate Gaussian on $(\bb, w_1, \ldots w_N),$ truncated to $y_i w_i \geq 0$ for $i=1,\ldots, N$.
The untruncated Gaussian has zero mean and precision matrix 
\eqan
\bM &= &
\pmat
M_{\bb \, \bb} & M_{\bb \, w}
\\
M_{w \, \bb}  & M_{ww} 
\pman  \qquad \in \mathbb{R}^{p+N, p+N}
\\
&=& \pmat
\sigma^{-2}\mathbb{I}_p + B B^T & B 
\\
B^T & \mathbb{I}_N
\pman 
\label{mbb}
\enan
where
\eqan
B^T  &=&  
\pmat
\bz_1
\\
\vdots
\\
\bz_N
\pman
\qquad \in \mathbb{R}^{N,p}
\enan
% \eqan
% M_{ww} &=& \mathbb{I}_{N}  \qquad \in \mathbb{R}^{N,N}
% \\
% M_{w \, \bb} &=&  M_{\bb \, w}^T= 
% \pmat
% \bz_1
% \\
% \vdots
% \\
% \bz_N
% \pman
% \qquad \in \mathbb{R}^{N,p}
% \\
% M_{\bb \, \bb} &=& \sigma^{-2} \mathbb{I}_p + \sum_{i=1}^{N} \bz_i \bz_i^T \qquad \in \mathbb{R}^{p,p}
% \label{mbb}
% \enan
We can sample from the posterior in  (\ref{post}) by sampling from the truncated Gaussian for $(\bb, w_1, \ldots w_N)$ and keeping only the~$\bb$ values.
It is easy to show that without the term $\sigma^{-2}\mathbb{I}_p$ in (\ref{mbb}), coming from the prior $p(\bb),$  the precision matrix would have $p$ null directions and our method would
not be applicable, since we assume the precision matrix to be positive definite. 
Note that the dimension of the TMG grows linearly with the number $N$ of data points. 
\begin{figure}[t]
\begin{center}
\includegraphics[scale=1]{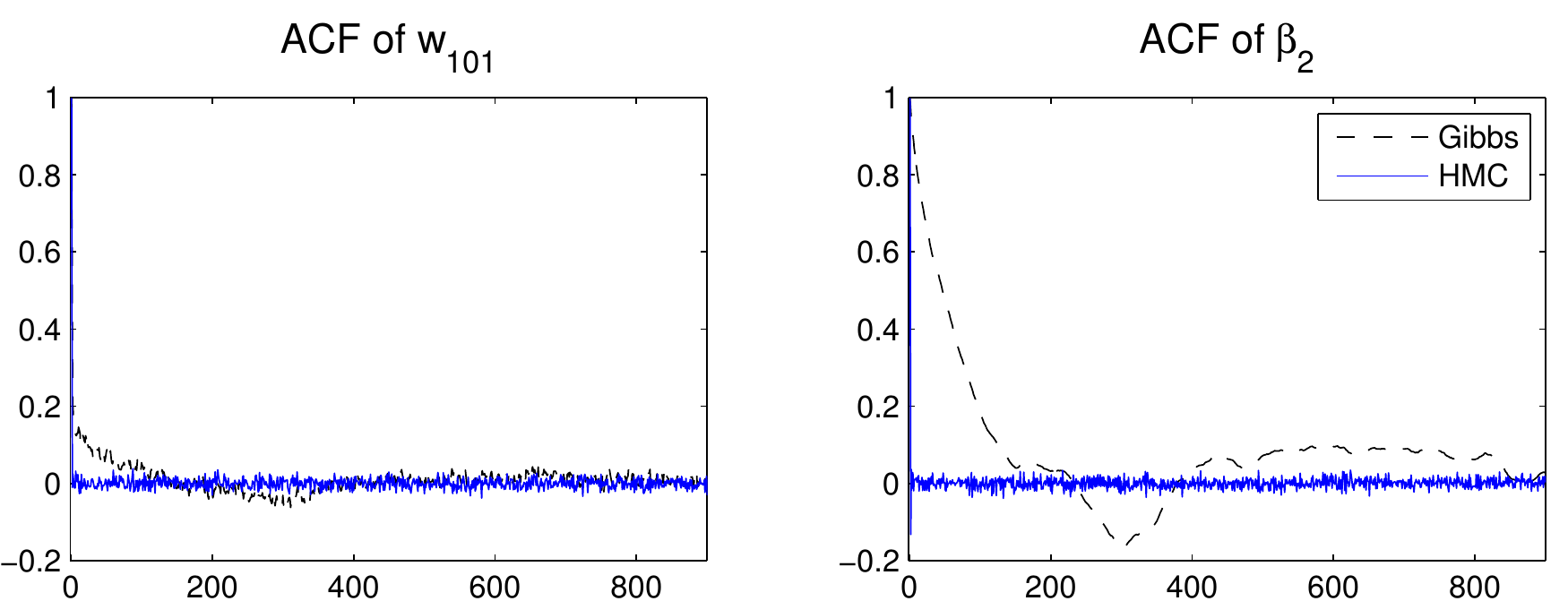}
\end{center}
\caption{{\bf Autocorrelation functions.} 
\emph{Autocorrelation functions (ACFs) for the first 900 lags of two of the 803 variables sampled in the Probit example.  
See Table~\ref{table_ess} for a comparison of efficiency measures of these two variables between the two samplers.
}}
\label{acfs_fig}
\end{figure}

The  structure of (\ref{mbb}) leads to a simple form for the untruncated covariance  $\Sigma= \bM^{-1}$,
\eqan 
\bM^{-1} &=&
\pmat 
\mathbb{I}_p & 0 
\\
-B^T & \mathbb{I}_N
\pman 
\pmat 
\sigma^2 \mathbb{I}_p & 0 
\\
0 & \mathbb{I}_N
\pman 
\pmat 
\mathbb{I}_p & -B
\\
0 & \mathbb{I}_N
\pman
\\
&=& Z^T Z
\enan 
where
\eqan 
Z^T = 
\pmat 
\mathbb{I}_p & 0 
\\
-B^T & \mathbb{I}_N
\pman 
\pmat 
\sigma \mathbb{I}_p & 0 
\\
0 & \mathbb{I}_N
\pman \,,
\enan 
and this form is such that acting with $Z^T$ takes $O(N)$ time, instead of $O(N^2)$ .

As an illustration,  
Figure~\ref{probit_fig} shows the values of $\bb$, sampled using Gibbs and exact HMC, from the posterior of a model with $p=3$ where $N=800$ data points were generated.
We used $z^1_i=1$, $z^2_i \sim Unif[-5,+5]$ and $z^3_i \sim {\cal N }(-4,\sigma=4)$. 
The values of $y_i$ were generated with $\beta_1=-9, \beta_2=20, \beta_3=27$ and we assumed a Gaussian prior with $\sigma^2=1$.  
Note that the means of the sampled $\beta_i$'s are different from the $\beta_i$'s  used to generate the data, due to the prior which pulls the $\beta_i$'s 
towards zero.

A more quantitative comparison between the HMC and the Gibbs samplers is presented in Figure~\ref{acfs_fig}, which compares the autocorrelation functions (ACFs) for two of the 
variables of the Probit model, and in Table~\ref{table_ess}, which compares, for the same variables, the Effective Sample Factors (ESFs) 
and the Effective Sample Size (ESS) normalized by CPU time. 
For the latter, we see a remarkable difference of two to three orders of magnitude between the two samplers.
\begin{table}[t]
\begin{center}
\begin{tabular}{|l|c|l|c|c|}
  \hline
   & Avg. CPU time   & & $w_{101}$ & $\beta_2$
  \\
  \hline \hline 
  HMC & $235.5$ secs.  & ESF  & {\bf 1.96 (1.28/7.03)  }  & { \bf 2.65 (1.28/3.69)  } \\
  \cline{3-5}
       &   & ESS/CPU  & {\bf 49.9 (32.6/178.6) } & {\bf 67.69 (32.6/92.84) } \\
  \hline \hline
  
  Gibbs & $651.7$ secs. &  ESF  & 0.037 (0.027/0.054) & 0.0051 (0.0036/0.0068) \\
    \cline{3-5}
  & & ESS/CPU  & 0.34 (0.22/0.50)   & 0.047 (0.033/0.063) \\  
  \hline
\end{tabular}
\end{center}
\caption{{\bf Efficiency of HMC vs. Gibbs in the Probit example.}  
\emph{The results summarize 10 runs of each sampler, and  
 correspond to the means of the variables $w_{101}$ and $\beta_2$ from the $803$-dimensional Probit example. 
The autocorrelation functions of these variables are shown in Figure~\ref{acfs_fig}. 
In each run we used 6,000 samples, after discarding 2,000 samples as burn-in.
Again, the CPU time for each run is random, but its variability across runs is negligible due to the large number of samples.
For ESF and  ESS/CPU, we show median (first/third quartiles).
ESS/CPU is the Effective Sample Size in units of the CPU runtime. 
For Gibbs we used the algorithm  of~\citep{damien2001sampling}. 
Note that the efficiency criteria are fairly variable across runs,
but the HMC ESS/CPU is consistently two to three  orders of magnitude bigger
than Gibbs.
The Gibbs sampler shows a big difference in the efficiency between the two variables (cf. their ACFs in Figure~\ref{acfs_fig}),
while this difference is minor in the HMC case.
The HMC runtime can be reduced further by parallelizing the computation of the travel times to hit each wall. }}
\label{table_ess}
\end{table}

A model related to the Probit is the Tobit model for censored data~\citep{tobin1958estimation}, which is a linear regression model 
where negative values are not observed:
\eqan 
y_i  = \left\{
 \begin{array}{ll}
y_i^* &  \textrm{for} \, y_i^*  > 0 \,,
\\
  0    & \textrm{for} \, y_i^*  \leq 0 \,,
\end{array}
\right.  
\enan 
where
\eqan
y_i^*= \bz_i \cdot \bb + \varepsilon_i\,,  \qquad \varepsilon_i \sim {\cal N}(0,\sigma) .
\enan
The likelihood of a pair $(y_i,\bz_i)$  is
\eqan 
p(y_i|\bz_i, \bb, \sigma) = \left\{
 \begin{array}{ll}
\frac{e^{-\frac{(y_i-\bz_i \cdot \bb)^2}{2 \sigma^2}}}{\sqrt{2 \pi \sigma^2}} &  \textrm{for} \, y_i  > 0 \,,
\\
\frac{1}{\sqrt{2 \pi \sigma^2}}  \int_{-\infty}^0 dw_i e^{-\frac{(w_i-\bz_i \cdot \bb)^2}{2 \sigma^2}}   & \textrm{for} \, y_i  = 0 \,,
\end{array}
\right.  
\enan 
and the posterior probability for $\bb$ is
\eqan
p(\bb| \textbf{Y}, \textbf{Z}, \sigma) &\vp& p(\bb| \sigma)\,  \prod_{i=1}^{N} p(y_i|\bz_i, \bb, \sigma)
\\
&\vp&  p(\bb | \sigma) \,  \prod_{i, y_i>0} e^{-\frac{(y_i- \bz_i \cdot \bb)^2}{2 \sigma^2}} 
 \prod_{i, y_i=0} \int_{-\infty}^0 \!\!  dw_i \,\, e^{-\frac{(w_i- \bz_i \cdot \bb)^2}{2 \sigma^2}}  
\enan
As in (\ref{post}), this can be treated as a  marginal distribution over the variables $w_i$, with the joint distribution for  $(\bb,w_i)$   a truncated multivariate Gaussian.

One can similarly consider multivariate Probit~\citep{ashford1970multi} and Tobit models. 
For the former, the Bayesian approach has been shown to be superior to Maximum Likelihood~\citep{geweke1994alternative}.

\subsection{Sample paths in a Brownian bridge}
\label{brbridge}
\begin{figure}[t]
\begin{center}
\includegraphics[scale=.95]{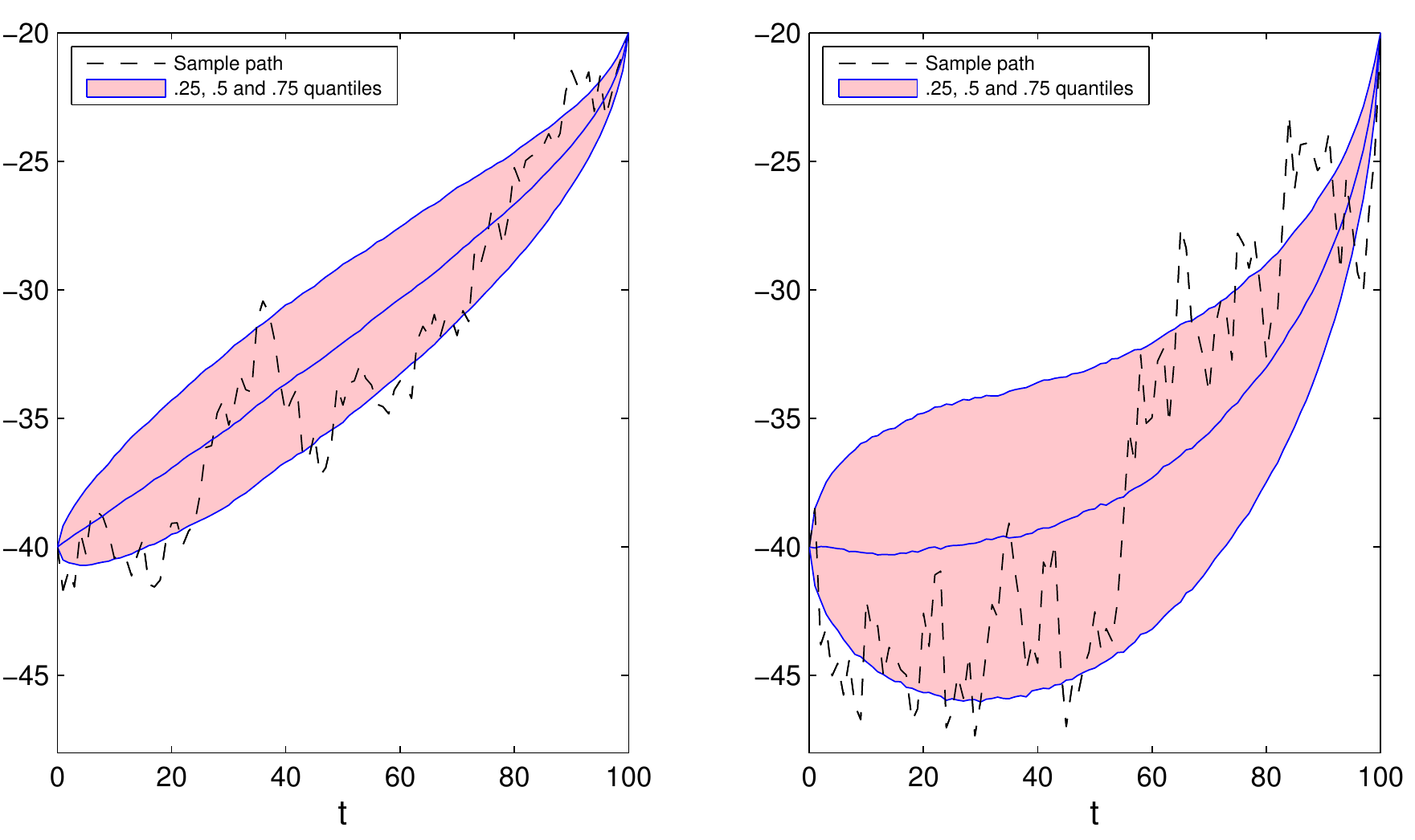}
\end{center}
\caption{{\bf Sample paths in a Brownian bridge.} \emph{Each figure shows one sample path and the median and $.25$ and $.75$ quantiles at each $t$
from $15,000$ sample paths (after 500 burn-in paths)
of a Brownian bridge model with $L=-40,$ $H=-20$ and $T=100$. 
Left: low noise, $\sigma^2=1$. Right: high noise, $\sigma^2=5$. 
%The tridiagonal form~(\ref{Mtri}) of the precision matrix $\bM$ reduces the runtime of the division by the Cholesky factor of $\bM$, needed for each sample, from $O(T^2)$ to $O(T)$.
} }
\label{bridge_fig}
\end{figure}

A family of cases where a simple special structure for $\bM$ arises are 
linear-Gaussian state-space models with constraints on the hidden variables. 
For concreteness, we will study the case of the Brownian bridge. 
Consider the following discrete stochastic process
\eqan 
V_t &=& V_{t-1} + \varepsilon_t \qquad \varepsilon_t \sim {\cal N}(0,\sigma^2)     \qquad t =1, \ldots, T
\\
V_0 &=& L \leq H
\\
V_t & < & V_T  \quad \quad \quad \quad \quad  \textrm{for} \,\,\,\,\,  t<T
\\
V_T &=& H
\enan 
The starting point $V_0 =L$ is fixed and the number of steps $T$  until the first hit of $H$ is a random variable.
This process is called the Brownian bridge and has applications, among others, in finance (see e.g.~\citep{glasserman2003monte} ) and in neuroscience, 
where it corresponds to a stochastic integrate-and-fire model 
(see e.g. \citep{paninski2004maximum}). Given $T$, we are interested in samples from the $(T-1)$-dimensional space of possible paths.
The log-density is given by
\eqan 
\log p(V_1, \ldots, V_{T-1}|V_0, V_T,\sigma^2) &=& -\frac{1}{2\sigma^2} \sum_{t=1}^{T} (V_t- V_{t-1})^2 + const.
\\
&=& 
-\frac12 V^T \bM V + \br^TV + const.
\enan 
with 
\eqan
V_t < H \qquad \qquad \textrm{for} \qquad t =1, \ldots T-1, 
\enan 
and we defined
\eqan 
V^T  &=& (V_1, \ldots,  V_{T-1})  \qquad \in \mathbb{R}^{T-1}
\enan
\eqan 
\bM &=& \sigma^{-2} 
\pmat
2 &  -1 		\\
-1   &   2  & -1 		\\
   &       &  . &   &  & 	    \\
   &       &    &  . &  & 	    \\
   &       &  -1 &  2 & -1 & 	    \\
   &       &     & -1 & -2 & -1	    \\
   &       &     &    &  -1  & 2	
\pman  \quad \in \mathbb{R}^{T-1 \times T-1}
\label{Mtri}
\\
\br^T  &=& (\sigma^{-2} V_0, 0, \ldots, 0, \sigma^{-2} V_T)  \qquad \in \mathbb{R}^{T-1}
\enan 
This is a $(T-1)$-dimensional TMG with  tridiagonal  precision matrix $\bM$. Therefore, in the Cholesky decomposition $\bM=R^TR$, $R$ is bidiagonal
and  the action of $R^{-1}$ takes $O(T)$ time instead of $O(T^2)$. Figure~\ref{bridge_fig} shows samples from this distribution.

\subsection{Bayesian splines for positive functions}
\label{splines}
Suppose we have noisy samples $(y_i,x_i), \, i=1\ldots N,$ from an unknown smooth positive function $f(x) >0,$ with $x  \in [0,h]$.
We can estimate $f(x)$  using cubic splines with knots at the $x_i$'s, plus $0$ and $h$~\citep{green1994nonparametric}.
The dimension of the vector space of cubic splines with $N$ inner knots is $N+4$.
Our model is thus
\eqan
y_i = \sum_{s=1}^{N+4}  a_s\phi_s(x_i) + \varepsilon_i  \quad \varepsilon_i \sim {\cal N}(0,\sigma)      \qquad i=1 \ldots N \,,
\label{smodel}
\enan 
where the functions $\phi_s(x)$ are a spline basis.
Suppose we are interested in the value of $f(x)$ at the points $x=z_j$ with $j=1\ldots m$.
To enforce  $f(x) >0$ at those points, we impose the constraints
\eqan
\bphi(z_j)\cdot \ba  \geq 0 \,, \qquad j=1\ldots m\,,
\label{phidota}
\enan
where
\eqan 
\bphi (x) &=& (\phi_1 (x), \ldots, \phi_{N+4} (x)) \,,
\\
\ba &=& (a_1, \ldots, a_{N+4}) \,.
\enan 
To obtain a  sparse constraint matrix, it is convenient to use the  B-spline basis, in which only four elements in the vector $\bphi(z_j)$ are non-zero 
for any $j$ (see, e.g. \citep{de2001practical} for details).  In a Bayesian approach, we are interested in sampling from the posterior distribution
\eqan
p(\ba, \sigma^2| \bY, \bX, \lambda) \vp p(\bY|\bX, \ba, \sigma^2) p(\ba|\lambda, \sigma^2) p(\sigma^2) \,,
\label{asigma_post}
\enan 
where we defined 
\eqan 
\bY &=& (y_1, \ldots, y_N) \,,
\\
\bX &=& (x_1, \ldots, x_N) \,.
\enan 
\begin{figure}[t]
\begin{center}
\includegraphics[scale=.85]{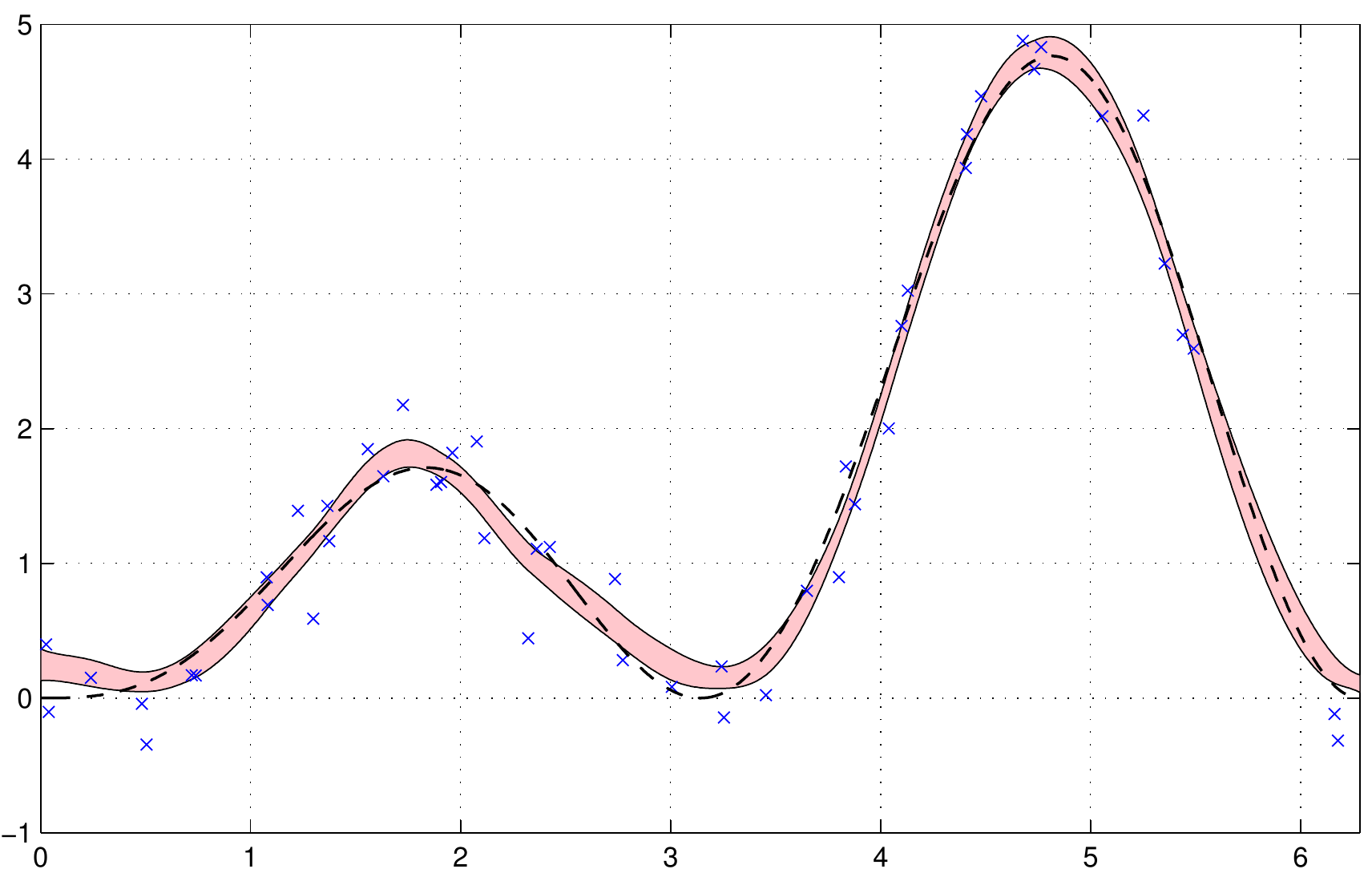}
\end{center}
\caption{{\bf Bayesian splines for positive functions.} 
\emph{The crosses show $50$   samples from $y_i = x_i\, \sin^2(x_i) + \varepsilon_i$, where $\varepsilon_i \sim {\cal N}(0,\sigma^2)$ with $\sigma^2=.09$. 
The values of  $x_i$ were sampled uniformly from $[0, 2\pi]$. The curve $f(x) = x\, \sin^2(x)$ is shown as a dashed line.
The shaded band shows   the splines  built with coefficients from the $.25$ and $.75$ quantiles of samples from the posterior distribution of $\ba$ in (\ref{asigma_post}).
We used a Jeffreys prior for $\sigma^2$~\citep{jeffreys} and imposed the positivity constraints~(\ref{phidota}) at $100$ points spread uniformly in $[0,2\pi]$.
The smoothness parameter~$\lambda$ was estimated  as $\hat{\lambda}= 0.0067$ by maximizing the marginal likelihood (empirical Bayes criterion), 
using a Monte Carlo EM algorithm. The mean of the samples of $\sigma^2$ was $\hat{\sigma^2} = 0.091$. 
The spline computations were performed with the ``fda'' MATLAB package~\citep{ramsay2009functional}.
}}
\label{splines_fig}
\end{figure}
The likelihood is 
\eqan 
p(\bY|\bX, \ba, \sigma^2) = 
\frac{1}{({2 \pi \sigma^2})^{N/2}} \exp \left( -\frac{1}{2 \sigma^2} \sum_{i=1}^{N} (y_i - \ba \cdot \bphi(x_i)  )^2 \right) \,,
\enan 
and for the prior on $\ba$  we consider 
\eqan 
% p(\ba, \sigma^2 | \lambda)  = p(\ba| \sigma^2 , \lambda) p( \sigma^2 )
% p( \sigma^2 ) &=& \frac{1}{\sigma^2} \,,
% \\
p(\ba| \lambda, \sigma^2) &\vp& \left( \frac{\lambda}{\sigma^2} \right)^{\frac{N+4}{2}} 
\exp \left( - \frac{\lambda}{2 \sigma^2} \int_{0}^{h} dx \, \left( \ba \cdot \bphi''(x) \right)^2 \right) \,,
\label{roughness}
\\
&\vp& \left( \frac{\lambda}{\sigma^2} \right)^{\frac{N+4}{2}} \exp \left( - \frac{\lambda}{2 \sigma^2} \, \ba^T \bK \ba \right) \,,
\label{roughness2}
\enan
where $\bK \in \mathbb{R}^{N+4, N+4}$ has entries 
\eqan  
K_{sr} = \int_{0}^{h} dx \, \phi''_s(x) \phi''_r(x) \,.
\enan 
The prior (\ref{roughness})-(\ref{roughness2}) is standard in the spline literature and imposes a $\lambda$-dependent penalty on the roughness of the estimated polynomial, 
with a bigger $\lambda$ corresponding to a smoother solution.  This  penalty helps to avoid  overfitting the data~\citep{green1994nonparametric}.
%As we show below, $\lambda$ can be estimated using an empirical Bayes criterion.

We can Gibbs sample from the posterior (\ref{asigma_post}) by alternating between the conditional distributions of $\sigma^2$ and $\ba$. 
% The former is 
% an inverse gamma distribution,
% \eqan 
% p(\sigma^2|\ba,  \bY, \bZ, \lambda) = \frac{b^a}{\Gamma (a)} \frac{1}{\sigma^{2 +2a}} \exp\left( -\frac{b}{\sigma^2} \right) \,,
% \label{invgamma}
% \enan 
% with
% \eqan
% a &=& N+2 \,,
% \\
% b &=& \frac{1}{2 } \sum_{i=1}^{N} (y_i - \ba \cdot \bphi(x_i)  )^2 + \frac{\lambda}{2 } \, \ba^T \bK \ba  \,,
% \enan
The latter is  a TMG with
\eqan
\log p(\ba | \sigma^2 , \bX, \bY, \lambda ) \vp - \frac{1}{2 \sigma^2} \, \ba^T (\bM+ \lambda \bK)  \ba + \frac{1}{\sigma^2}\ba^T \cdot \br
\,,   \qquad s=1\ldots N+4\,,
\label{tmg_splines}
\enan
constrained by (\ref{phidota}), and we defined 
\eqan 
\bM &=& \sum_{i=1}^N \bphi(x_i) \bphi(x_i)^T  \qquad \in \mathbb{R}^{N+4,N+4} \,,
\\
\br &=&  \sum_{i=1}^N y_i \bphi(x_i) \qquad \in \mathbb{R}^{N+4} \,.
\enan 
In the B-spline basis, the matrices $\bM$ 
and $\bK$ in (\ref{tmg_splines}) have a banded form~\citep{de2001practical}.
As in Example~\ref{brbridge} above, this allows us to speed up the runtime from $O(N^2)$ to $O(N)$ for each sample.

% \noindent
% In a Bayesian setting, the parameter $\lambda$ can be chosen  as the value that maximizes the integrated posterior,
% \eqan 
% \hat{\lambda} = \arg \max_{\substack{\lambda}} \int \! d\ba \, d \sigma^2 p(\ba, \sigma^2 | \bX, \bY, \lambda ) \,.
% \enan 
% This value  can be estimated using the EM algorithm: given an approximate solution $\lambda^{(k)}$, the next estimation is given by
% \eqan 
% \lambda^{(k+1)} &=& \arg \max_{\substack{\lambda}} \int \! d\ba \, d \sigma^2 \log p(\ba, \sigma^2 | \bX, \bY, \lambda ) p(\ba, \sigma^2 | \bX, \bY, \lambda^{(k)} ) \,,
% \\ 
% & \simeq& \arg \max_{\substack{\lambda}} \, \frac{1}{J}\sum_{j=1}^{J} \log p(\ba_{j|k}, \sigma^2_{j|k} | \bX, \bY, \lambda ) \,,
% \\
% &=& \frac{(N+4)J}{\sum_{j=1}^{J}  \sigma^{-2}_{j|k}\,\, \ba_{j|k}^T \bK \ba_{j|k}       }  \,,
% \label{lest}
% \enan 
% where $\ba_{j|k}$ and $\sigma^{2}_{j|k}$   are $J$ samples from (\ref{asigma_post}) with  $\lambda=\lambda^{(k)}$.
% We iterate the updates (\ref{lest}) until convergence.
% 

\noindent
Figure \ref{splines_fig} shows an example for the function $f(x) = x\, \sin^2(x)$,
with
$N=50$ points sampled as
\eqan
y_i = x_i\, \sin^2(x_i) + \varepsilon_i   \quad \varepsilon_i \sim {\cal N}(0,\sigma) \qquad \sigma^2=.09\,,
\label{splinesamples}
\enan 
and with the $x_i$ sampled uniformly from $[0, 2\pi]$. 

% The shaded band shows  the splines built with coefficients $\ba$ from the $.25$ and $.75$ quantiles of samples from the posterior distribution (\ref{asigma_post}).
% We estimated $\lambda$  using the EM algorithm (\ref{lest}) as $\hat{\lambda}= 0.0073$, and the mean of the samples of $\sigma^2$ was $\hat{\sigma^2} = 0.0891$. 

\subsection{Bayesian reconstruction of quantized stationary Gaussian processes}
\begin{figure}[t]
\begin{center}
\includegraphics[scale=.95]{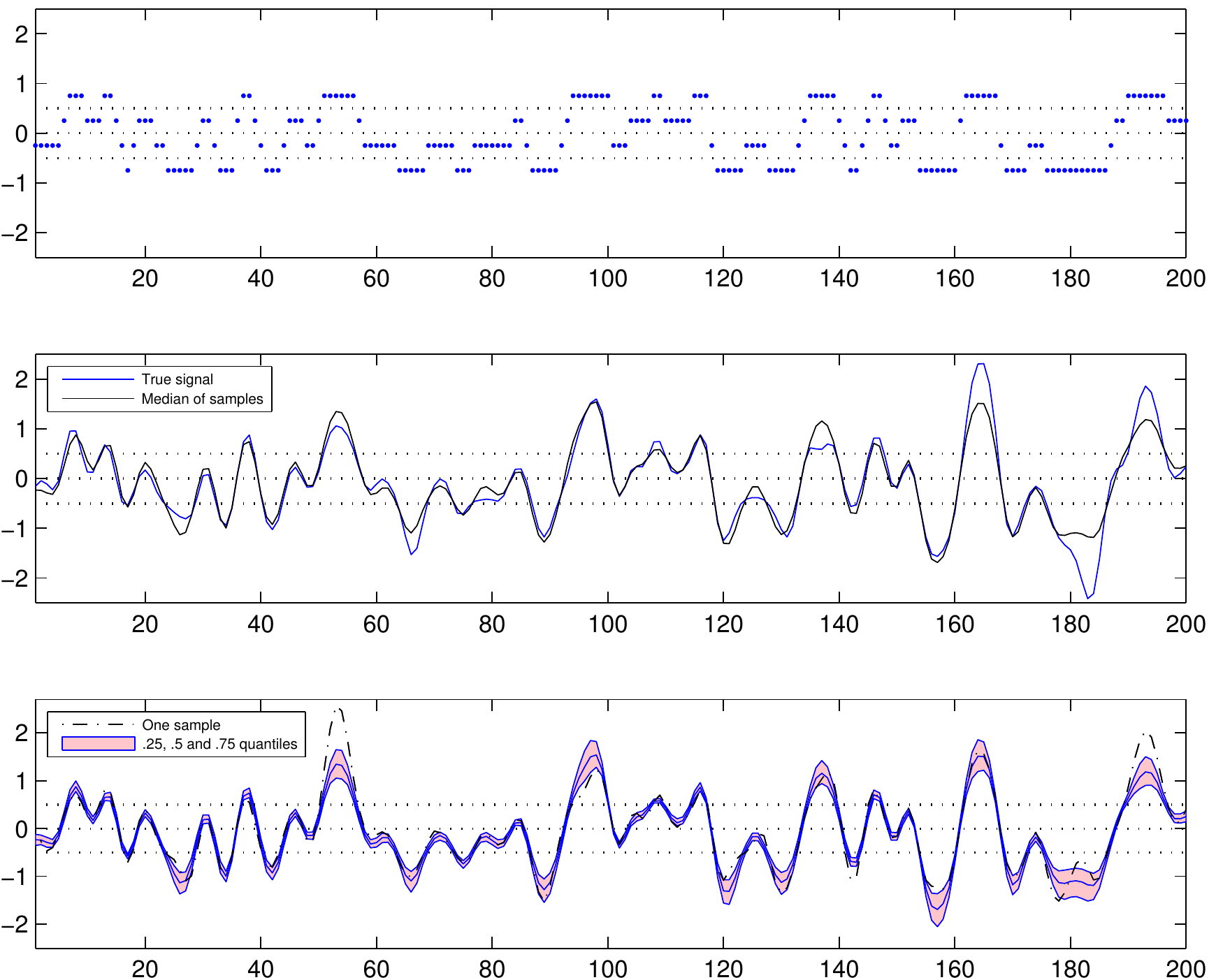}
\end{center}
\caption{{\bf Bayesian reconstruction of quantized functions.} \emph{Upper panel: a  function quantized
at  $N=200$ points, taking four possible values. Middle panel: true function and median of the posterior samples.
Lower panel: one posterior sample and median and $.25$ and $.75$ quantiles of the posterior samples.  
The three dotted lines, at $y= 0, \, \pm .5$, separate the four quantization regions.
We used $5,000$ samples, after discarding $500$ as burn-in. The $N=200$ values of the true function were sampled from a Gaussian process with kernel 
given by (\ref{kernel}) with $\sigma^2 = 0.6, \, \eta^2 =0.2$. 
Using the translation invariance of this kernel, the covariance matrix of the Gaussian can be embedded into a circulant matrix,
and this allows  to sample the initial velocity at every iteration of the HMC sampler in $O(N \log N) $ time instead of $O(N^2)$.
} }
\label{quantized_fig}
\end{figure}
Suppose that we are given $N$  values of a function  $f(x_i) \,, \, i=1,\ldots, N$,  that takes discrete values in a set 
$\{q_k\}\,, \, k=1,\ldots, K $. We assume that this is a quantized projection 
of a sample $y(x_i)$ from a stationary Gaussian process with a known translation-invariant  covariance kernel of the form 
\eqan
(M^{-1})_{ij} = \Sigma_{ij} = K(|x_i-x_j|)\,,
\label{qcov}
\enan 
and the quantization follows a known rule of the form
\eqan
f(x_i) = q_k  \qquad \textrm{if} \qquad z_k \leq y(x_i) < z_{k+1} \,.
\label{qtrunc}
\enan 
We are interested in sampling the posterior distribution 
\eqan
p(y(x_1), \ldots , y(x_N)| f(x_1), \ldots, f(x_N))  \,,
\enan 
which is a Gaussian with covariance (\ref{qcov}), truncated by the quantization rules (\ref{qtrunc}).
In this case, we can exploit the Toeplitz form of the $N$-by-$N$-covariance matrix $\Sigma_{ij}$ 
to reduce the runtime of the HMC sampler. 
The idea is to embed $\Sigma$ into a  circulant matrix as (see e.g.,~\citep{chu1999inside}) 
\eqan
C&=&
\pmat
\Sigma_{11} &  \Sigma_{12} &.  &.  & \Sigma_{1(N-1)} & \Sigma_{1N} & \Sigma_{1(N-1)} & . & . & \Sigma_{12} 		            \\
\Sigma_{12} &  \Sigma_{11} &.  &.  & \Sigma_{1(N-2)} & \Sigma_{1(N-1)} & \Sigma_{1N} & . & . & \Sigma_{13} 		            \\
 & & & & . & .\\
 & & & & . &. \\
\pman  \quad \in  \mathbb{R}^{(2N-2) \times (2N-2)}
%\\
%&\equiv& 
%\pmat \Sigma & U \\
%U^T & V 
%\pman
\enan
As is well known, a circulant matrix is diagonalized by the discrete Fourier transform matrix $O$ as
\eqan 
C = O^T \Lambda O \,,
\enan 
where $O^T O=1$ and $\Lambda$ is the diagonal matrix of the eigenvalues.
We can exploit this structure in the frame in which the Hamiltonian is given by (\ref{hamnw}). Recall that at each iteration we must sample
the initial velocity $\dot{X}(0)$ from an N-dimensional Gaussian with covariance $\Sigma$. We can do this by starting with  a 
$(2N-2)$-dimensional sample $\varepsilon$ from a unit-covariance Gaussian and computing 
\eqan 
%\pmat 
%\dot{X}(0)
%\\
%\tilde{X}(0) \pman
%= 
O^T 
\Lambda^{1/2}  \varepsilon \,.
\label{multo}
\enan 
The first $N$ elements of this vector are samples from a Gaussian with covariance $\Sigma$ as desired.
% \eqan 
% \langle \dot{X}(0) \dot{X}^T(0) \rangle = \Sigma \,,
% \enan 
The multiplication by $O^T$ in (\ref{multo}) can be performed with a Fast Fourier Transform, and this  reduces the runtime for each $\dot{X}(0)$ sample 
from $O(N^2)$ to $O(N \log N)$. Figure~\ref{quantized_fig} shows an example of a quantized and reconstructed function evaluated at $N=200$ points. The function was sampled from 
a Gaussian process with kernel
\eqan 
K(|x_i-x_j|) = \sigma^2 e^{-\frac{|x_i-x_j|^2}{2\eta^2}}
\label{kernel}
\enan 
with $\sigma^2 = 0.6, \, \eta^2 =0.2$, and the quantization was performed by splitting the target space into four regions 
separated by $y= 0, \, \pm .5$. 

This technique  can be easily extended to functions defined in higher dimensional spaces, such as images in 2D, where the quantization acts as a lossy compression technique
and the estimated values $E(y(x_i)| f(x_1), \ldots, f(x_N))$  can be thought of as the decoded (decompressed) image.

\section{The Bayesian Lasso}
\label{baylasso}
The techniques introduced above can also be used to sample from
multivariate distributions whose log density is piecewise quadratic,
with linear or elliptical boundaries between the piecewise
regions. Instead of presenting the most general case, let us elaborate
the details for the example of the Bayesian Lasso
\citep{park2008bayesian,hans2009bayesian,polson2011bayesian}.

We are interested in the posterior distribution of the coefficients $\bb \in \mathbb{R}^d$ and $\sigma^2$  of a linear regression model
\eqan
y_n = \bz_n \cdot \bb + \varepsilon_n \qquad \qquad \varepsilon_n \sim {\cal N}(0,\sigma^2)
\qquad \qquad n=1, \ldots, N\,,
\enan
Defining
\eqan 
\bY &=& (y_1, \ldots, y_N)
\\
\bZ &=& (\bz_1, \ldots, \bz_N) \,,
\enan 
we want to sample from the posterior distribution
\eqan
p(\bb, \sigma^2| \bY, \bZ, \lambda) \vp p(\bY|\bZ, \bb, \sigma^2) p(\bb|\lambda, \sigma^2) p(\sigma^2) \,,
\label{beta_sigma_post}
\enan 
with prior density for the coefficients
\eqan
% p(\sigma^2) &=& \frac{1}{\sigma^2}
% \\
p(\bb|\lambda, \sigma^2) &=& \left( \frac{\lambda}{2\sigma^2} \right)^{d} \exp \left( -\frac{\lambda}{\sigma^2} \sum_{i=1}^{d} |\beta_i| \right)\,.
\label{beta_prior}
\enan
This prior  is called the {\it lasso} (for `least absolute shrinkage and selection operator') 
and imposes a $\lambda$-dependent sparsening penalty  in the maximum likelihood solutions for $\bb$~\citep{Tibs96}.

We can Gibbs sample from the posterior (\ref{beta_sigma_post}) by alternating between the conditional distributions of~$\sigma^2$ and~$\bb$. 
% The former is 
% given an  inverse gamma distribution as in (\ref{invgamma}), with
% \eqan
% a &=& \frac{N}{2} + d
% \\
% b &=& \frac12 \sum_{n=1}^{N} (y_n - \bz_n \cdot \bb)^2 + \lambda \sum_{i=1}^{d} |\beta_i| \,.
% \enan
The latter is given by  
\eqan
-\log p(\bb|\bY, \bZ,  \sigma^2, \lambda) &=& \frac{1}{2 \sigma^2} \sum_{n=1}^{N} (\bz_n \cdot \bb - y_n)^2 + \frac{\lambda}{ \sigma^2} \sum_{i=1}^{d} |\beta_i| + const.
\label{beta_p}
\\
&=& \frac{1}{2 \sigma^2} \bb^T \bM \bb - \frac{1}{ \sigma^2}\sum_{i=1}^{d} L^i(s_i) \beta_i + const. 
\label{beta_post}
\enan
where we defined
\eqan
\bM &=&  \sum_{n=1}^N \bz_n \bz_n^T \qquad \in \mathbb{R}^{d \times d}
\\
L^i(s_i) &=& \sum_{n=1}^{N}y_n(z_i)_n  - \lambda s_i   \qquad i = 1, \ldots, d.
\enan
with
\eqan
s_i = \textrm{sign}(\beta_i) \,.
\enan 
Sampling $\bb$ from (\ref{beta_post}) was considered  previously via Gibbs sampling, either 
expressing the Laplace prior~(\ref{beta_prior}) as mixtures of Gaussians~\citep{park2008bayesian}
or Bartlett-Fejer kernels~\citep{polson2011bayesian}, or directly from (\ref{beta_post})~\citep{hans2009bayesian}.

% \begin{figure}[t]
% \begin{center}
% \includegraphics[scale=.95]{bayesian_lasso.pdf}
% \end{center}
% \caption{{\bf Bayesian Lasso.}  \emph{First 200 samples of the HMC Bayesian Lasso sampler applied to the $N=442, d=10$ Diabetes dataset 
% described in~\citep{efr}. Each curve corresponds to samples of each of the ten $\beta_i$ variables of the dataset. 
% The value of $\lambda$ was 
% estimated by maximizing the marginal likelihood, using a Monte Carlo EM algorithm.
% The efficiency of our  algorithm is comparable to other methods 
% such as~\citep{park2008bayesian}, but our approach allows to easily incorporate constraints on the $\beta_i$ variables.
% }}
% \label{bayesian_lasso_fig}
% \end{figure}
% 

In order to apply Hamiltonian Monte Carlo  we consider the Hamiltonian
\eqan
H = \frac{1}{2 \sigma^2} \bb^T \bM \bb - \frac{1}{ \sigma^2} \sum_{i=1}^{d} L^i(s_i) \beta_i + \frac{\sigma^2}{2 } \bP^T \bM^{-1} \bP \,.
\label{hamb}
\enan
Note that we did not map the coordinates to a canonical frame, as in Section~\ref{Linear}. Instead, we chose a momenta mass matrix 
 $\sigma^{-2}\bM$, which is equal to the precision matrix of the coordinates. This choice leads to 
the simple equations
% \eqan
% \dot{\beta}_i &=& \frac{\partial H}{\partial p^i} = \sigma^2 \sum_{j=1}^{d} M^{-1}_{i j}p^{j}
% \\
% \dot{p}^i &=& -\frac{\partial H}{\partial \beta_i} = - \frac{1}{ \sigma^2} \sum_{j=1}^{d} M^{i j}\beta_{j} + \frac{1}{ \sigma^2} L^i(s_i)
% \enan
%which can be combined to 
\eqan
\ddot{\beta}_i = - \beta_i + \mu_i(\bs) \,,
\label{eomb}
\enan
where 
\eqan
\mu_i(\bs) = \sum_{j=1}^{d} M_{i j}^{-1}L^j(s_j) \,.
\label{mbs}
\enan
The solution to  (\ref{eomb}) is
\eqan
\label{esolb}
\beta_i(t) &=& \mu_i(\bs) + a_i \sin(t) + b_i \cos(t) \,,
\\
&=& \mu_i(\bs) + A_i \cos(t + \varphi_i) \,,
\label{sol2}
\enan
where
\eqan
A_i &=& \sqrt{a_i^2 + b_i^2} \,,
\\
\tan \varphi_i &=&  -\frac{a_i}{b_i}  \,.
\enan
The constants $a_i, b_i$ in (\ref{esolb}) can be expressed in terms of the initial conditions as
\eqan
b_i &=& \beta_i(0) - \mu_i(\bs)
\label{bb}
\\
a_i &=& \dot{\beta}_i(0)
\label{abz}
\\
&=& M^{-1}_{i j } p^j(0).
\label{ab}
\enan
As in Section~\ref{Linear}, we start by sampling $\bP$ from $p(\bP) = {\cal N}(0,\sigma^{-2} \bM)$ and 
let the particle  move during a time $T=\pi/2$.
The trajectory of the particle is given by (\ref{sol2}) until a coordinate crosses any of the~$\beta_i=0$ planes, which happens at the smallest time  $t>0$  such that 
\eqan
0 = \mu_i(\bs) + A_i \cos(t + \varphi_i) \,, \qquad \qquad i=1,\ldots, d.
\label{lasso_cond}
\enan 
(Note that
had we transformed the coordinates $\bb$ to a canonical frame, each
condition here would have involved a sum of $d$
terms; thus the parameterization we use here leads to sparser, and
therefore faster, computations.)
Suppose the constraint is met for $i=j$ at time $t=t_j$. At this point
$\beta_j$ changes sign, so the Hamiltonian (\ref{hamb}) changes by
replacing \eqan L^j(s_j) \longrightarrow L^j(-s_j) = L^j(s_j) + 2s_j
\lambda \,, \enan which in turn changes the values of $\mu_i(\bs)$'s
in (\ref{mbs}). Note from (\ref{eomb}) that this causes a jump in
$\ddot{\bb}(t_j)$.  Using the continuity
of~$\bb(t_j)$,~$\dot{\bb}(t_j)$ and the updated~$\mu_i(\bs)$'s, we can
compute new values for $a_i$ and $b_i$ as in (\ref{bb}) and
(\ref{abz})~to continue the trajectory for times $t>t_j$.  

%Figure~\ref{bayesian_lasso_fig} illustrates the result of using this algorithm in the Diabetes dataset of~\citep{efr}.

We have found the efficiency of this algorithm comparable to other methods
to sample from the Bayesian Lasso model, e.g.,~\citep{park2008bayesian}.
The real advantage of our approach would be  when the coefficients $\beta_i$'s have additional constraints,
as in the tree shrinkage
model~\citep{leblanc1998monotone}, the hierarchical
Lasso~\citep{bien2012lasso}, or when some of the coefficients are constrained to be positive.
In these cases, it is very easy to combine this  algorithm with the
imposition of constraints of the previous Section.

Finally, the piecewise linear log-density (\ref{beta_p}) is continuous with
discontinuous derivative, but we can also consider discontinuous
log-densities defined piecewise.  In these cases, the velocity is not
continuous across the boundary of two regions, but jumps in such a way
that the total energy is conserved.  The extension of the basic method
to this case is straightforward.

\section*{Acknowledgements}
This work was supported by an NSF CAREER grant, a McKnight Scholar
award,  NSF grant IIS-0904353 and by the Defense Advanced
Research Projects Agency (DARPA) MTO under the auspices of Dr. Jack
Judy, through the Space and Naval Warfare Systems Center, Pacific
Grant/Contract No. N66001-11-1-4205.
This material is based upon work supported by, or in part by, the U.
S. Army Research Laboratory and the U. S. Army Research Office under
contract number W911NF-12-1-0594.
AP is supported by the Swartz
Foundation.  We thank Matt Hoffman, Alexandro Ramirez, Carl Smith and Eftychios Pnevmatikakis for helpful discussions.

\newpage 
\bibliographystyle{plainnat}
%\bibliographystyle{myagsm}
%\bibliography{mybib,jv-paper,thebib,rafa2}
%\bibliography{thebib}
\bibliography{/home/aripakman/Desktop/my_papers/2012/synaptic/latex/thebib}

\begin{thebibliography}{41}
\providecommand{\natexlab}[1]{#1}
\providecommand{\url}[1]{\texttt{#1}}
\expandafter\ifx\csname urlstyle\endcsname\relax
  \providecommand{\doi}[1]{doi: #1}\else
  \providecommand{\doi}{doi: \begingroup \urlstyle{rm}\Url}\fi

\bibitem[Albert and Chib(1993)]{albert1993bayesian}
J.H. Albert and S.~Chib.
\newblock Bayesian analysis of binary and polychotomous response data.
\newblock \emph{Journal of the American statistical Association}, pages
  669--679, 1993.

\bibitem[Ashford and Sowden(1970)]{ashford1970multi}
JR~Ashford and RR~Sowden.
\newblock Multi-variate probit analysis.
\newblock \emph{Biometrics}, pages 535--546, 1970.

\bibitem[Bannerman et~al.(2011)Bannerman, Sargant, and Lue]{JCC:JCC21915}
M.~N. Bannerman, R.~Sargant, and L.~Lue.
\newblock Dynamo: a free o(n) general event-driven molecular dynamics
  simulator.
\newblock \emph{Journal of Computational Chemistry}, 32\penalty0 (15):\penalty0
  3329--3338, 2011.

\bibitem[Bien et~al.(2012)Bien, Taylor, and Tibshirani]{bien2012lasso}
J.~Bien, J.~Taylor, and R.~Tibshirani.
\newblock {A Lasso for Hierarchical Interactions}.
\newblock \emph{Arxiv preprint arXiv:1205.5050}, 2012.

\bibitem[Chen and Deely(1992)]{chen1992application}
M.H. Chen and J.~Deely.
\newblock {Application of a new Gibbs Hit-and-Run sampler to a constrained
  linear multiple regression problem}.
\newblock Technical report, Technical Report 92-21, Purdue University, Center
  for Statistical Decision Sciences and Department of Statistics, 1992.

\bibitem[Chu and George(1999)]{chu1999inside}
E.~Chu and A.~George.
\newblock \emph{Inside the FFT black box: serial and parallel fast Fourier
  transform algorithms}.
\newblock CRC, 1999.

\bibitem[Cox and Wermuth(2002)]{cox2002some}
DR~Cox and N.~Wermuth.
\newblock {On some models for multivariate binary variables parallel in
  complexity with the multivariate Gaussian distribution}.
\newblock \emph{Biometrika}, 89\penalty0 (2):\penalty0 462--469, 2002.

\bibitem[Damien and Walker(2001)]{damien2001sampling}
P.~Damien and S.G. Walker.
\newblock Sampling truncated normal, beta, and gamma densities.
\newblock \emph{Journal of Computational and Graphical Statistics}, 10\penalty0
  (2):\penalty0 206--215, 2001.

\bibitem[De~Boor(2001)]{de2001practical}
C.~De~Boor.
\newblock \emph{A practical guide to splines}.
\newblock Springer Verlag, 2001.

\bibitem[Duane et~al.(1987)Duane, Kennedy, Pendleton, and
  Roweth]{duane1987hybrid}
S.~Duane, A.D. Kennedy, B.J. Pendleton, and D.~Roweth.
\newblock Hybrid monte carlo.
\newblock \emph{Physics letters B}, 195\penalty0 (2):\penalty0 216--222, 1987.

\bibitem[Ellis and Maitra(2007)]{ellis2007multivariate}
N.~Ellis and R.~Maitra.
\newblock {Multivariate Gaussian simulation outside arbitrary ellipsoids}.
\newblock \emph{Journal of Computational and Graphical Statistics}, 16\penalty0
  (3):\penalty0 692--708, 2007.

\bibitem[Emrich and Piedmonte(1991)]{emrich1991method}
Lawrence~J Emrich and Marion~R Piedmonte.
\newblock A method for generating high-dimensional multivariate binary
  variates.
\newblock \emph{The American Statistician}, 45\penalty0 (4):\penalty0 302--304,
  1991.

\bibitem[Gelfand et~al.(1992)Gelfand, Smith, and Lee]{gelfand1992bayesian}
Alan~E Gelfand, Adrian~FM Smith, and Tai-Ming Lee.
\newblock Bayesian analysis of constrained parameter and truncated data
  problems using gibbs sampling.
\newblock \emph{Journal of the American Statistical Association}, 87\penalty0
  (418):\penalty0 523--532, 1992.

\bibitem[Gelman et~al.(2004)Gelman, Carlin, Stern, and
  Rubin]{gelman2004bayesian}
A.~Gelman, J.B. Carlin, H.S. Stern, and D.B. Rubin.
\newblock \emph{Bayesian data analysis}.
\newblock CRC press, 2004.

\bibitem[Geweke(1991)]{geweke1991efficient}
J.~Geweke.
\newblock Efficient simulation from the multivariate normal and student-t
  distributions subject to linear constraints and the evaluation of constraint
  probabilities.
\newblock In \emph{Computing Science and Statistics: Proceedings of the 23rd
  Symposium on the Interface}, pages 571--578, 1991.

\bibitem[Geweke et~al.(1994)Geweke, Keane, and Runkle]{geweke1994alternative}
J.~Geweke, M.~Keane, and D.~Runkle.
\newblock Alternative computational approaches to inference in the multinomial
  probit model.
\newblock \emph{The review of economics and statistics}, pages 609--632, 1994.

\bibitem[Glasserman(2003)]{glasserman2003monte}
P.~Glasserman.
\newblock \emph{Monte Carlo methods in financial engineering}, volume~53.
\newblock Springer, 2003.

\bibitem[Green and Silverman(1994)]{green1994nonparametric}
P.J. Green and B.W. Silverman.
\newblock \emph{Nonparametric regression and generalized linear models: a
  roughness penalty approach}, volume~58.
\newblock Chapman \& Hall/CRC, 1994.

\bibitem[Hammersley and Morton(1956)]{hammersley1956new}
JM~Hammersley and KW~Morton.
\newblock {A new Monte Carlo technique: antithetic variates}.
\newblock In \emph{Mathematical Proceedings of the Cambridge Philosophical
  Society}, volume~52, pages 449--475. Cambridge Univ Press, 1956.

\bibitem[Hans(2009)]{hans2009bayesian}
C.~Hans.
\newblock Bayesian lasso regression.
\newblock \emph{Biometrika}, 96\penalty0 (4):\penalty0 835--845, 2009.

\bibitem[Herbison-Evans(1994)]{herbison1994solving}
D.~Herbison-Evans.
\newblock Solving quartics and cubics for graphics.
\newblock Technical report, Technical Report TR-94-487, Basser Department of
  Computer Science, University of Sidney, Sidney, Australia, 1994.

\bibitem[Hoffman and Gelman(2011)]{hoffman2011no}
M.D. Hoffman and A.~Gelman.
\newblock {The No-U-Turn sampler: adaptively setting path lengths in
  Hamiltonian Monte Carlo}.
\newblock \emph{Arxiv preprint arXiv:1111.4246}, 2011.

\bibitem[Izaguirre and Hampton(2004)]{izaguirre2004shadow}
J.A. Izaguirre and S.S. Hampton.
\newblock {Shadow hybrid Monte Carlo: an efficient propagator in phase space of
  macromolecules}.
\newblock \emph{Journal of Computational Physics}, 200\penalty0 (2):\penalty0
  581--604, 2004.

\bibitem[Jeffreys(1946)]{jeffreys}
Harold Jeffreys.
\newblock An invariant form for the prior probability in estimation problems.
\newblock \emph{Proceedings of the Royal Society of London. Series A,
  Mathematical and Physical Sciences}, 186\penalty0 (1007):\penalty0 pp.
  453--461, 1946.
\newblock ISSN 00804630.
\newblock URL \url{http://www.jstor.org/stable/97883}.

\bibitem[Kennedy(1990)]{kennedy1990theory}
AD~Kennedy.
\newblock The theory of hybrid stochastic algorithms.
\newblock In \emph{NATO ASIB Proc. 224: Probabilistic Methods in Quantum Field
  Theory and Quantum Gravity}, volume~1, page 209, 1990.

\bibitem[Kennedy and Bitar(1994)]{kennedy1994exact}
AD~Kennedy and KM~Bitar.
\newblock {An exact Local Hybrid Monte Carlo algorithm for gauge theories}.
\newblock \emph{Nuclear Physics B-Proceedings Supplements}, 34:\penalty0
  786--788, 1994.

\bibitem[Kotecha and Djuric(1999)]{kotecha1999gibbs}
J.H. Kotecha and P.M. Djuric.
\newblock Gibbs sampling approach for generation of truncated multivariate
  gaussian random variables.
\newblock In \emph{Proceedings., 1999 IEEE International Conference on
  Acoustics, Speech, and Signal Processing}, volume~3, pages 1757--1760. IEEE,
  1999.

\bibitem[LeBlanc and Tibshirani(1998)]{leblanc1998monotone}
M.~LeBlanc and R.~Tibshirani.
\newblock Monotone shrinkage of trees.
\newblock \emph{Journal of Computational and Graphical Statistics}, pages
  417--433, 1998.

\bibitem[Liu(2008)]{liumonte}
J.S. Liu.
\newblock \emph{Monte Carlo strategies in scientific computing}.
\newblock Springer, 2008.

\bibitem[Neal(2010)]{neal2010mcmc}
R.M. Neal.
\newblock {MCMC using Hamiltonian dynamics}.
\newblock \emph{Handbook of Markov Chain Monte Carlo}, 54:\penalty0 113--162,
  2010.

\bibitem[Neelon and Dunson(2004)]{neelon2004bayesian}
B.~Neelon and D.B. Dunson.
\newblock Bayesian isotonic regression and trend analysis.
\newblock \emph{Biometrics}, 60\penalty0 (2):\penalty0 398--406, 2004.

\bibitem[Paninski et~al.(2004)Paninski, Pillow, and
  Simoncelli]{paninski2004maximum}
L.~Paninski, J.W. Pillow, and E.P. Simoncelli.
\newblock Maximum likelihood estimation of a stochastic integrate-and-fire
  neural encoding model.
\newblock \emph{Neural computation}, 16\penalty0 (12):\penalty0 2533--2561,
  2004.

\bibitem[Park and Casella(2008)]{park2008bayesian}
T.~Park and G.~Casella.
\newblock {The Bayesian lasso}.
\newblock \emph{Journal of the American Statistical Association}, 103\penalty0
  (482):\penalty0 681--686, 2008.

\bibitem[Polson and Scott(2011)]{polson2011bayesian}
N.G. Polson and J.G. Scott.
\newblock {The Bayesian Bridge}.
\newblock \emph{Arxiv preprint arXiv:1109.2279}, 2011.

\bibitem[Ramsay et~al.(2009)Ramsay, Hooker, and Graves]{ramsay2009functional}
J.O. Ramsay, G.~Hooker, and S.~Graves.
\newblock \emph{Functional data analysis with R and MATLAB}.
\newblock Springer Verlag, 2009.

\bibitem[Rasmussen(2003)]{rasmussen2003gaussian}
C.E. Rasmussen.
\newblock {Gaussian processes to speed up Hybrid Monte Carlo for expensive
  Bayesian integrals}.
\newblock In \emph{Bayesian Statistics 7: Proceedings of the 7th Valencia
  International Meeting}, pages 651--659. Oxford University Press, 2003.

\bibitem[Robert(1995)]{robert1995simulation}
C.P. Robert.
\newblock Simulation of truncated normal variables.
\newblock \emph{Statistics and computing}, 5\penalty0 (2):\penalty0 121--125,
  1995.

\bibitem[Robert and Casella(2004)]{robert2004monte}
C.P. Robert and G.~Casella.
\newblock \emph{Monte Carlo statistical methods}.
\newblock Springer Verlag, 2004.

\bibitem[Rodriguez-Yam et~al.(2004)Rodriguez-Yam, Davis, and
  Scharf]{rodriguez2004efficient}
G.~Rodriguez-Yam, R.A. Davis, and L.L. Scharf.
\newblock {Efficient Gibbs sampling of truncated multivariate normal with
  application to constrained linear regression}.
\newblock \emph{Unpublished Manuscript}, 2004.
\newblock \url{ http://www.stat.columbia.edu/~rdavis/papers/CLR.pdf}.

\bibitem[Tibshirani(1996)]{Tibs96}
R.~Tibshirani.
\newblock Regression shrinkage and selection via the lasso.
\newblock \emph{Journal of the Royal Statistical Society. Series B},
  58:\penalty0 267--288, 1996.

\bibitem[Tobin(1958)]{tobin1958estimation}
J.~Tobin.
\newblock Estimation of relationships for limited dependent variables.
\newblock \emph{Econometrica: Journal of the Econometric Society}, 26\penalty0
  (1):\penalty0 24--36, 1958.

\end{thebibliography}
\end{document}